\documentclass[twoside,letterpaper,aps,prb,floatfix,twocolumn,showpacs,superscriptaddress,10pt,nobibnotes]{revtex4-1}
\usepackage{graphicx,times}
\usepackage{amssymb,amsmath}
\usepackage{multirow}
\usepackage{color}
\usepackage{epstopdf}
\usepackage{psfrag}
\usepackage[colorlinks=true, urlcolor=blue, linkcolor=red, citecolor=blue, pdftex]{hyperref}
\hypersetup{pdfstartview={FitV}, pdfpagemode = UseNone, pdfpagelayout=SinglePage, bookmarksnumbered={true}}

\begin{document}

\title{Combined analytical and numerical approach to magnetization 
 plateaux in one-dimensional spin tube antiferromagnets}

\author{X. Plat}
\affiliation{Laboratoire de Physique Th\'eorique,  IRSAMC, CNRS and Universit\'e de Toulouse, UPS, F-31062 Toulouse, France}

\author{S. Capponi}
\affiliation{Laboratoire de Physique Th\'eorique,  IRSAMC, CNRS and Universit\'e de Toulouse, UPS, F-31062 Toulouse, France}

\author{P.\ Pujol}
\affiliation{Laboratoire de Physique Th\'eorique,  IRSAMC, CNRS and Universit\'e de Toulouse, UPS, F-31062 Toulouse, France}

\date{\today}


\begin{abstract}
In this paper, we investigate the properties of frustrated three-leg spin tubes 
under a magnetic field. We concentrate on two kind of geometries for these 
tubes, one of which is relevant for the compound $\mathrm{[(CuCl_2tachH)_3Cl]Cl_2}$. 
We combine an analytical path integral approach with a strong coupling approach, as well as large-scale 
 Density Matrix Renormalization Groups (DMRG) simulations, to identify the 
presence of plateaux in the magnetization curve as a function of the value of 
spin $S$. We also investigate the issue of gapless non-magnetic excitations on 
some plateaux, dubbed chirality degrees of freedom for both tubes. 
\end{abstract}

\maketitle

\section{Introduction}\label{sec:intro}
Nowadays one-dimensional and quasi one-dimensional antiferromagnetic (AF)
quantum spin systems are a very active theme in condensed matter
physics. Thanks to efforts in chemical synthesis, it is now possible to obtain
materials which can be effectively considered as one-dimensional
systems, making possible to verify the theoretical
predictions. 

The natural extension of quantum spins chains are quantum spin ladders, 
which are made of two or more coupled chains. These ladders represent the first step between one and
two-dimensional systems. They give rise to interesting features~\cite{Dagotto1996} and have 
been extensively studied over the last decades. So far, various properties have 
been established both analytically and experimentally.~\cite{Schulz1986,White1994,Senechal1995,Cabra1997,Cabra1998,Kojima1995,Eccleston1994}
Going back to the problem of a single Heisenberg spin-$S$ chain, 
we know since the work of Haldane~\cite{Haldane1983} that the chain
is gapless (respectively gapped) if $S$ is half-integer (respectively integer). 
In a similar way, there is a parity effect of the number of coupled half-integer
spin chains to form the ladder.~\cite{Strong1992} A gap opens in the spectrum of
antiferromagnetic spin ladders with an even number of legs and their
spin correlation functions decay exponentially. On the other hand, for 
an odd number of legs of half-integer spin, such ladders have massless excitations above
their ground state and the decay of the spin correlations is
algebraic. Experimental investigations have confirmed these
predictions, for example the observation of a gap in the spin-$1/2$
two-leg ladders $\mathrm{SrCu_2O_3}$~\cite{Azuma1994} and $\mathrm{Cu_2(C_5H_{12}N_2)_2Cl_4}$.~\cite{Chaboussant1997}

New properties arise when the role of transverse boudary counditions is taken
into account. The results quoted above are valid only for ladders,
which correspond to open boundary condition (OBC). Applying periodic boundary 
conditions (PBC) in the rung direction to form a
spin tube seems to cause the opening of a gap in both even and odd
cases. For an even number of legs, this is explained in terms of the formation
of spin spinglets in the transverse direction. The reason is different
in the odd case, where the PBC induce geometrically frustrated
interactions. The direct consequence of this frustration is a twofold 
degenerate dimerized ground state with an excitation gap above it.~\cite{Kawano1997}

Among the questions emerging from the study of these quasi one-dimensional systems, an
important one concerns their magnetization process when an external
magnetic field is turned on. Classicaly, the magnetization curve of such
systems is expected to be a straight line until the saturation field. 
But at low enough temperatures, quantum effects begin to play a role and 
magnetization plateaux can appear. This has been observed in various chains and 
ladders spin systems.~\cite{Narumi1998,Shiramura1998,Yoshida2005,Kikuchi2005}

A condition, neither sufficient nor necessary, for the existence of magnetization plateaux in a
quantum spin-$S$ chain has been found by Oshikawa, Yamanaka and
Affleck.~\cite{Oshikawa1997} This condition, which was later extended to ladder 
systems,~\cite{Cabra1998} restricts the possible values of
the magnetization for a plateau. It reads
\begin{equation}
N(S-m)\in\mathbb{Z}
\label{eq:OYA_condition},
\end{equation}
where $m$ is the on-site magnetization and $N$ the number of spins per
unit cell. This result has been obtained through a generalization of
the Lieb-Shultz-Mattis (LSM) theorem.~\cite{Lieb1961} Although the low-energy state
constructed in this approach has the same total magnetization as the
ground state, they present bosonization arguments indicating that, in
general, low-energy states appear also in different magnetization
sector. In 2009, Tanaka, Totsuka and Hu (TTH) have used a spin
coherent-states path integral approach to recover this
condition.~\cite{Tanaka2009} The main advantage of their method is
that it can be applied for any value $S$.

In this paper we investigate successively the effect of a magnetic
field on two different types of three-leg spin tube of spin-$S$,
namely the simple spin tube and the twisted spin tube, which will be
describe below.  For each one, we combine analytical and numerical
methods and proceed as follows. First we apply to the spin tube the
TTH approach, which take into account the effects of the Berry phase
appearing in the partition function.  This leads to an effective field
theory for the spin tube, from which a condition on the magnetization
plateaux is infered and we give an estimate the region of existence of the
plateaux. We also discuss the question of having gapless non-magnetic
excitations but also several different gapped phases for these excitations. 
Then, in the case of half-integer spins we study the 
limit of strongly coupled chains. A new non-magnetic degree of freedom
appears in this regime for the magnetizations of the lowest and
highest magnetization plateaux, namely a right or left chirality. It
comes from the twofold degeneracy of the ground state. In this limit
we show the possibility of the existence on those plateaux of a
quantum phase transition. As the coupling along the chains is
increased, the chirality degree of freedom may go from a critical to a
gapped regime. This behaviour has recently been observed in the simple
spin tube of spin-1/2 by Okunishi {\it et al.} using density-matrix
renormalization-group (DMRG) calculations.~\cite{Okunishi2012} We
expect this transition to be well accounted by a pertubated XXZ effective
Hamiltonian and support this statement by a very simple qualitative
numerical result. Then we perform DMRG calulations to study 
the magnetization process. Analyzing the entanglement entropy and the local 
magnetizations, we finally examine the different phases occuring for 
the chirality on the plateaux.

The rest of the paper is organized as follows. In
Sec.~\ref{sec:simple_spintube}, we consider the case of the simple three-leg spin
tube by using the path-integral approach to understand the appearance
of some magnetization plateaux, then we investigate the role of the
chirality degrees of freedom, and finally we compare our predictions to 
DMRG simulations in the $S=3/2$ case.
In Sec.~\ref{sec:twisted_spintube}, the case of the twisted spin tube
is addressed following the same strategy. Finally, we draw some conclusions and 
discuss possible perspectives in Sec.~\ref{sec:conclusion}. Some technical 
details about the duality transformation are given in appendix.


\section{Simple spin tube}\label{sec:simple_spintube}
\subsection{The model}\label{sec:simple_spintube_model}
\begin{figure}[!htb]
\includegraphics[width=0.26\textwidth,clip]{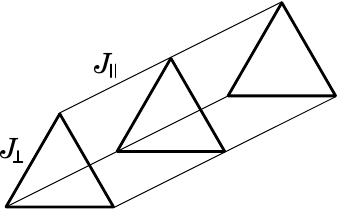}
\caption{Lattice structure of the simple spin tube.}
\label{fig:simple_spintube_picture}
\end{figure}

The Hamiltonian of the simple spin tube, which is the first geometry that we 
consider, reads
\begin{equation}
\begin{split}
&H=H_{\perp}+H_{\parallel}+H_{h}\\
&H_{\perp}=J_{\perp}\sum_j\sum_{\alpha=1,2,3}\vec{S}_{\alpha,j}.\vec{S}_{\alpha+1,j}\\
&H_{\parallel}=J_{\parallel}\sum_j\sum_{\alpha=1,2,3}\vec{S}_{\alpha,j}.\vec{S}_{\alpha,j+1}\\
&H_h=-h\sum_j\sum_{\alpha=1,2,3}S_{\alpha,j}^z,
\end{split}
\label{eq:simple_spintube_Hamiltonian}
\end{equation}
where $\vec{S}_{\alpha,j}$ is the spin-$S$ operator, $J_{\parallel}>0$
is the intrachain AF coupling, $J_{\perp}>0$ the AF rung coupling and
$h$ the magnetic field along the $z$ axis (Fig.~\ref{fig:simple_spintube_picture}). 
The subscript $i$ ($\alpha$) represents the site number in the chain (rung)
direction. The tube structure (PBC in the rung direction) induces
frustration in this simple nearest-neighbour model. Applied to this
model, the OYA condition (\ref{eq:OYA_condition}) takes the form
\begin{equation}
3(S-m)\in\mathbb{Z}
\label{eq:OYA_simple_spintube_condition},
\end{equation}
as there are three spins per unit cell.

So far, the Hamiltonian (\ref{eq:simple_spintube_Hamiltonian}) has
already been investigated in previous works. By using bosonization
techniques for the spin $S=1/2$ and in zero magnetic field, Schulz has
suggested that the tube has a spin gap induced by the geometric
frustration.~\cite{Schulz1997} DMRG calculations have confirmed this
statement by establishing the existence of a spin
gap,~\cite{Kawano1997} which is explained in terms of a two-fold
degenerate dimerized ground-state with broken translation
symmetry. New questions arise when one of the couplings between the rungs
is changed, making the tube asymmetric.  In that case, it seems that
the spin gap vanishes for a small but finite
asymmetry.~\cite{Sakai2008} The effect of a magnetic field has also be
addressed.~\cite{Tandon1999,Orignac2000,Citro2000,Sato2007,Sato2007b}
For higher half-integer spins, only a few results are available. In
the $S=3/2$ case, the existence of a spin gap, as for $S=1/2$, has
been reported recently.~\cite{Nishimoto2011} Moving to integer spins
$S$, the simple spin tube displays interesting properties. For weakly
coupled chains, another parity effect has been established for the
low-energy properties using the Non-Linear Sigma Model (NLSM). For an
odd number of legs, the lowest magnon band of the model
(\ref{eq:simple_spintube_Hamiltonian}) is six-fold degenerate, compared
to the three-fold degeneracy of non-frustrated systems, namely
 tubes with an even number of legs or ladders.~\cite{Sato2005,Sato2005a,Sato2007a}
Turning on a uniform or non-uniform magnetic field, Sato predicted a succession of 
quantum phase transitions between critical phases as the field is
increased, along with again an even-odd
effect.~\cite{Sato2005} Introducing an asymmetry in the rung
couplings, a NLSM analysis and DMRG results have shown the existence
of $2S$ quantum phase transitions between gapped phases when varying
the anisotropy parameter.~\cite{Charrier2010}

Finally, we discuss the \emph{classical configurations} of the spin tube
(\ref{eq:simple_spintube_Hamiltonian}). For decoupled triangles with no external field, 
the three spins are simply lying in the same plane with angles of $2\pi/3$ 
between them. If the triangles are now coupled by the $J_{\parallel}$ term, the situation does not change as 
the longitudinal coupling is not frustrating and satisfied with a $k_{\parallel}=\pi$ 
order in this direction. A magnetic field arrange the spins in an ``umbrella'' 
configuration, where the three spins on the triangle are equally
polarized and have angles of $2\pi/3$ between their projections in the
plane. Thus, the classical ground state of the simple spin tube 
is simply an umbrella configuration on each triangle with a canted
order along the tube. We parametrize it as
\begin{equation}
\vec{S}_{\alpha,j}=S
\begin{pmatrix} (-1)^j \mathrm{sin}(\theta_0)\mathrm{cos}(\varphi_{\alpha}^0)  \\ (-1)^j \mathrm{sin}(\theta_0)\mathrm{sin}(\varphi_{\alpha}^0) \\ \mathrm{cos}(\theta_0)
\end{pmatrix}
,
\label{eq:simple_spintube_classical_gs}
\end{equation}
where $\mathrm{cos}(\theta_0)=\frac{h}{S(3J_{\perp}+4J_{\parallel})}$
and $\varphi_{\alpha}^0=(\alpha-1)2\pi/3$.

This state breaks the U(1) symmetry around the $z$ axis, 
one site translations and parity transformations. More precisely, the latter one 
is a symmetry which is going to play a very important role in this system. It 
is related to what we dub the chirality degree of freedom. Consider the chirality vector order
parameter:
\begin{equation}
\chi_j=\frac{1}{3}\sum_{\alpha=1}^{3}(\vec{S}_{\alpha,j} \times \vec{S}_{\alpha+1,j})^z,
\label{eq:chirality_order_parameter}
\end{equation}
which is invariant under cyclic permutation of the three chain indices 
(i.e. translations in the transverse directions) but changes sign under the permutation
of two chains. For the classical configuration (\ref{eq:simple_spintube_classical_gs}), it reads 
$\chi_j \propto \sum_{\alpha} \mathrm{sin}((\varphi_{\alpha}^0-\varphi_{\alpha+1}^0)/2)$. 
Thus, the choice of a given classical configuration, namely the choice of 
$\varphi_{\alpha}^0-\varphi_{\alpha+1}^0=\pm 2\pi/3$, fixes the sign of this order 
parameter and breaks the $\mathbb{Z}_2$ symmetry explicitly. This in turn 
will have important consequences in the analysis of the chirality behaviour 
within the path integral approach.

\subsection{Path integral approach}\label{sec:simple_spintube_path}
\subsubsection{Derivation of a low-energy action}\label{sec:simple_spintube_derivation_action}

We begin the study of the simple spin tube (\ref{eq:simple_spintube_Hamiltonian})
following the method recently developped by Tanaka, Totsuka and Hu.~\cite{Tanaka2009} 
They used a Haldane's spin coherent-state~\cite{Klauder1979} path integral
approach to re-derive the OYA condition (\ref{eq:OYA_condition}) for the Heisenberg 
chain with easy-plane single-ion anisotropy. The interest of the method
is its validity for any value of the spin $S$. Haldane's
analysis leads to an action comprising two terms. One is the
coherent-state expectation value of the Hamiltonian, or simply the
Hamiltonian for the classical configuration. The other term is the Berry phase one and
corresponds to the surface area (or the solid angle), $\int \mathrm{d}\tau [1-\mathrm{cos}(\theta(\tau))]\partial_{\tau}\varphi(\tau)$ 
in spherical coordinates, enclosed by each spin during its imaginary-time 
$\tau$ evolution. We want to build a low-energy effective theory from this starting point. The 
method consists in finding the classical ground state of the system and 
then adding the quantum fluctuations to derive an effective action.

We start from the classical ground state discussed in
Sec.~\ref{sec:simple_spintube_model} and now we add the
fluctuations around this state, writing
\begin{equation}
\left\{
\begin{split}
&\theta_0 \rightarrow \theta_{\alpha,j}=\theta_0 + \delta \theta_{\alpha,j}\\ 
&\varphi_{\alpha}^0 \rightarrow \varphi_{\alpha}^0+\varphi_{\alpha,j}=(\alpha-1)\frac{2\pi}{3}+\varphi_{\alpha,j}
\end{split}
\right.
\label{eq:simple_spintube_flucuations}
\end{equation}
and expand the spin components up to second order in $\delta \theta$. 
The calculation of the SU(2) commutation relations $[S_{\alpha,i}^z,S_{\beta,j}^{\pm}]$ 
leads to introducing a new set of variables $\Pi_{\alpha,j}$, defined by
\begin{equation}
\Pi_{\alpha,j}=-S\left[\mathrm{sin}(\theta_0)\delta\theta_{\alpha,j}+\frac{1}{2}\mathrm{cos}(\theta_0)\delta\theta_{\alpha,j}^2\right],
\label{eq:simple_spintube_definition_pi_momentum}
\end{equation}
which are the conjugates of the $\varphi_{\alpha,j}$'s. It ensures to
have the correct commutators for the spin operators. Then we rewrite
these operators as functions of the conjugate fluctuations variables 
$\varphi_{\alpha,j}$ and $\Pi_{\alpha,j}$ as
\begin{widetext}
\begin{equation}
\left\{
\begin{split}
&S^{\pm}_{\alpha,j}\approx(-1)^je^{\pm i\left[(\alpha-1)\frac{2\pi}{3}+\varphi_{\alpha,j}\right]}S\left[\mathrm{sin}(\theta_0)-\frac{m}{S^2\mathrm{sin}(\theta_0)}\Pi_{\alpha,j}-\frac{1}{2}\frac{S^2}{S^2-m^2}\frac{1}{S^2\mathrm{sin}(\theta_0)}\Pi_{\alpha,j}^2\right]\\ 
&S^z_{\alpha,j}\approx S \mathrm{cos}(\theta_0)+\Pi_{\alpha,j}
\end{split}
\right.,
\label{eq:simple_spintube_spin_operators_pi}
\end{equation}
\end{widetext}
where $m=S\mathrm{cos}(\theta_0)$ is the classical magnetization per site.

Casting these expressions into the action, taking the continuum limit and
keeping terms up to second order in the fields, we obtain the
low-energy effective action
\begin{widetext}
\begin{equation}
\begin{split}
S[\{\Pi_{\alpha}\},\{\varphi_{\alpha}&\}]=\int d\tau dx \bigg\{\sum_{\alpha=1,2,3}\left[\frac{1}{2}aJ_{\parallel}(S^2-m^2)(\partial_x\varphi_{\alpha})^2 + a\left(2J_{\parallel}+\frac{1}{2}J_{\perp}\frac{S^2}{S^2-m^2}\right)\Pi_{\alpha}^2 \right] \bigg.\\
&+aJ_{\perp}\left(1-\frac{1}{2}\frac{m^2}{S^2-m^2}\right)(\Pi_1\Pi_2+\Pi_2\Pi_3+\Pi_3\Pi_1)
+\frac{J_{\perp}}{4}\frac{S^2-m^2}{a}\left[(\varphi_1-\varphi_2)^2+(\varphi_2-\varphi_3)^2+(\varphi_3-\varphi_1)^2\right]\\
&-\frac{\sqrt{3}}{2}mJ_{\perp}\left[\Pi_1(\varphi_3-\varphi_2)+\Pi_2(\varphi_1-\varphi_3)+\Pi_3(\varphi_2-\varphi_1)\right]\bigg. +i\sum_{\alpha=1,2,3}\left[\left(\frac{S-m}{a}\right)\partial_{\tau}\varphi_{\alpha}-\Pi_{\alpha}\partial_{\tau}\varphi_{\alpha}\right] \bigg\},
\end{split}
\label{eq:simple_spintube_low-energy_action}
\end{equation}
\end{widetext}
with $a$ the lattice constant. We see that all the fluctuations,
transerve or longitudinal, are coupled. The last two imaginary terms
come from the Berry phase part of the action. It is important to
stress that the $\partial_{\tau}\varphi_{\alpha}$ terms, although being
total derivatives, can not be dropped. Indeed, the fields
$\varphi_{\alpha}$ are angular variables defined on a circle and thus this term counts the
winding number of each field.

As at this order the action is gaussian in the fields $\Pi_{\alpha}$, we can integrate them out and the action becomes
\begin{widetext}
\begin{equation}
\begin{split}
S[\{\phi_{\alpha}\}]&=S_{ch}[\phi_1,\phi_2]+S_s[\phi_s]\\
S_{ch}[\phi_1,\phi_2]&=\int d\tau dx \bigg\{ \frac{1}{2}\lambda_{\tau}^{(1,2)}\left[(\partial_{\tau}\phi_1)^2+(\partial_{\tau}\phi_2)^2\right]+ \frac{1}{2}\lambda_x^{(1,2)}\left[(\partial_x\phi_1)^2+(\partial_x\phi_2)^2\right] +M^2(\phi_1^2+\phi_2^2) - i\mu(\phi_1\partial_{\tau}\phi_2-\phi_2\partial_{\tau}\phi_1)\bigg\}\\
S_s[\phi_s]&=\int d\tau dx \bigg\{ \frac{1}{2}\lambda_{\tau}^{(s)}(\partial_{\tau}\phi_s)^2+\frac{1}{2}\lambda_{x}^{(s)}(\partial_{x}\phi_s)^2+i3\frac{S-m}{a}\partial_{\tau}\phi_s \bigg\},
\end{split}
\label{eq:simple_spintube_lowenergy_action_pi_integrated}
\end{equation}
\end{widetext}
where $S_{ch}$ denotes, for reasons which will become clear later, the chirality 
part of the action and $S_s$ the symmetric one. We have made the change of 
variable $\vec{\phi}=U\vec{\varphi}$, where
\begin{equation}
\vec{\phi}=
\begin{pmatrix} \phi_1  \\ \phi_2 \\ \phi_s
\end{pmatrix}
,\ \ 
\vec{\varphi}=
\begin{pmatrix} \varphi_1  \\ \varphi_2 \\ \varphi_s
\end{pmatrix}
,\ \ 
U=
\begin{pmatrix} -\frac{1}{\sqrt{2}} & \frac{1}{\sqrt{2}} & 0 \\ -\frac{1}{\sqrt{6}} & -\frac{1}{\sqrt{6}} & \frac{2}{\sqrt{6}} \\ \frac{1}{\sqrt{3}} & \frac{1}{\sqrt{3}} & \frac{1}{\sqrt{3}} \end{pmatrix}
,
\label{eq:orthogonal_transformation_phi}
\end{equation}
and have rescaled the symmetric field as $\phi_s \rightarrow \phi_s/\sqrt{3}$. The coefficients of 
the action (\ref{eq:simple_spintube_lowenergy_action_pi_integrated}) read
\begin{equation}
\begin{split}
&\lambda_{\tau}^{(1,2)}=\frac{1}{a\left(4J_{\parallel}+\frac{3}{2}J_{\perp}\frac{m^2}{S^2-m^2}\right)},\ \ \lambda_{\tau}^{(s)}=\frac{3}{a(4J_{\parallel}+3J_{\perp})}, \\
&\lambda_{x}^{(1,2)}=aJ_{\parallel}(S^2-m^2),\ \ \lambda_{x}^{(s)}=3aJ_{\parallel}(S^2-m^2), \\
&M^2=3J_{\parallel}J_{\perp}\frac{S^2-m^2}{a\left(4J_{\parallel}+\frac{3}{2}J_{\perp}\frac{m^2}{S^2-m^2}\right)}, \\
&\mu=\frac{3}{2}J_{\perp}\frac{m}{a\left(4J_{\parallel}+\frac{3}{2}J_{\perp}\frac{m^2}{S^2-m^2}\right)}.
\end{split}
\label{eq:simple_spintube_constants}
\end{equation}

The symmetric field is now decoupled from $\phi_1$ and $\phi_2$, 
and we will study them separately.

\subsubsection{Symmetric action and magnetization plateaux}\label{sec:simple_spintube_symmetric_action}
We notice that the action $S_s$ for the $\phi_s$ field obtained in 
(\ref{eq:simple_spintube_lowenergy_action_pi_integrated}) has the same form than 
the action of the Heisenberg chain in a magnetic field.~\cite{Tanaka2009} The term 
$i\partial_{\tau}\phi_s$ comes directly from the Berry phase part of the action 
discussed above. In order to understand its role on the low-energy physics, we apply 
a duality transformation~\cite{Lee1991} on this action (details are given in Appendix A). 
The dual action finally reads
\begin{equation}
\begin{split}
S[\tilde{\Phi}_s]=\int d\tau dx \bigg\{& \frac{1}{2}K_s(\vec{\nabla}\tilde{\Phi}_s)^2 \bigg.\\
&+ g_1\mathrm{cos}\left(2\pi\left[\tilde{\Phi}_s+3\frac{S-m}{a}x \right]\right) \bigg\},
\end{split}
\label{eq:simple_spintube_dual_action}
\end{equation}
where $\tilde{\Phi}_s$ is the dual field, $K_s=1/\sqrt{\lambda_{\tau}^{(s)}\lambda_x^{(s)}}$, 
$\vec{\nabla}=(\partial_{\tau},\partial_x)$ and $g_1$ is a constant we have not computed.

We find a gaussian action perturbed by a cosine 
term. It is the sine-Gordon action in (1+1) dimension
plus a spatial modulation $2\pi 3(S-m)x/a$ of the cosine. This
modulation is a direct consequence of the topological Berry phase. We
can therefore separate two cases. If $3(S-m)\notin\mathbb{Z}$, the
cosine is incommensurate and will average to zero by integrating over
space. It remains a simple gapless gaussian model and there
will not be any plateau in the magnetization curve for general values of 
$m$. We can understand the effect of the Berry phase in terms of protecting 
the system from the vortices. On the other hand, if
\begin{equation}
3(S-m)\in\mathbb{Z}
\label{eq:simple_spintube_plateau_condition},
\end{equation}
the cosine is commensurate and we recover the sine-Gordon model. A
gap can open in the spectra, causing the emergence of plateaux in the
magnetization curve for these particular magnetizations. For
example, we expect to observe a plateau for average magnetization per
site $1/6, 1/2, 5/6$ and $7/6$ in the $S=3/2$ case.

More precisely, the presence of the cosine term is not sufficient to open
a gap. It has to be relevant in the renormalization-group (RG) sense and 
it will depend on the microscopic parameters. From the well-known action 
(\ref{eq:simple_spintube_dual_action}) (without modulation) we 
compute the scaling dimension
$\Delta $
\begin{equation}
\Delta=\pi\sqrt{\frac{3J_{\parallel}(S^2-m^2)}{J_{\perp}\left(1+\frac{4J_{\parallel}}{3J_{\perp}}\right)}}\ .
\label{eq:simple_spintube_cosine_scalingdim}
\end{equation}
For the cosine to be relevant $\Delta$ has to be smaller than $2$ and thus we expect plateaux in the weakly coupled triangles regime.

We end this discussion with some comments. First, it is worth 
noting that the condition (\ref{eq:simple_spintube_plateau_condition}) does 
not predict the $m=0$ plateau found numerically by DMRG.~\cite{Kawano1997} 
But in the action (\ref{eq:simple_spintube_dual_action}), 
higher harmonics of the cosine term have been dropped, especially the 
first one which actually predicts this plateau. However, as the 
harmonics terms would be less relevant than the fundamental, we 
can not conclude about the $m=0$ plateau with this analysis. 
In fact, the zero-field problem needs a different approach taking into 
account the O(3) symmetry of the model.~\cite{Charrier2010} 
Second, the action (\ref{eq:simple_spintube_dual_action}) 
is the same as the one obtained for spin-1/2 using bosonization~\cite{Oshikawa1997}, but
coming here from a large $S$ method. So, with the condition 
(\ref{eq:simple_spintube_plateau_condition}) we have simply 
recovered the OYA result (\ref{eq:OYA_simple_spintube_condition}) 
on the possible values for magnetization plateaux to occur. 

\begin{figure}[!htb]
\begin{center}
\includegraphics[width=0.45\textwidth,clip]{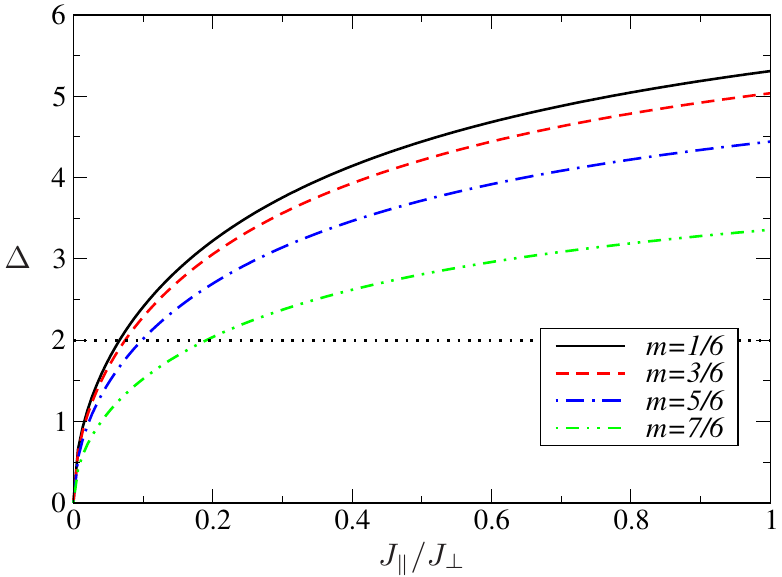}
\end{center}
\caption{(Color online) Scaling dimension of the cosine operator in the action 
(\ref{eq:simple_spintube_dual_action}) as a function of 
$J_{\parallel}/J_{\perp}$ for $S=3/2$ at the values of $m$ fulfilling the 
condition (\ref{eq:simple_spintube_plateau_condition}). The perturbation is relevant 
and opens a gap in the spectrum if $\Delta<2$, thus the plateaux should disappear 
for $J_{\parallel}/J_{\perp} \gtrsim 0.1$.}
\label{spin tube scaling dim S=3/2}
\end{figure}

Finally, it is important to note that the factor of $3$
in front of the Berry term $\frac{S-m}{a} \partial_{\tau}\phi_{s}$ is
not an artefact of the transformation (\ref{eq:orthogonal_transformation_phi}) 
and of the rescaling of the field. The rescaling is necessary for the following reason. 
As the $\vec{\varphi}$ fields are angular variable, they satisfy
\begin{equation}
\varphi_{i}(\tau=\beta)=\varphi_{i}(\tau=0)+2\pi n_{i},
\label{eq:simple_spintube_periodicity_condition}
\end{equation}
and the same condition for spatial periodic boundary conditions in $x=0$ and $x=L$. 
That gives similar conditions for the new fields $\vec{\phi}$, for instance the antisymmetric 
combination $\phi_1$ satisfies $\phi_1(\beta)=\phi_1(0)+2\pi(n_2-n_1)/\sqrt{2}$.
But, given the form of the action $S_{ch}$ for $\phi_1$ and $\phi_2$ (see below), we expect those 
fields to be small and therefore to have no winding, {\it i.e} $n_1=n_2=0$. It remains for $\phi_s$ 
the periodicity $\phi_s(\beta)=\phi_s(0)+2\pi n_3/\sqrt{3}$. Thus the rescaling 
$\phi_s \rightarrow \phi_s/\sqrt{3}$ is required to ensure a correct $2\pi$-periodicity.

\subsubsection{Chirality degree of freedom}\label{sec:simple_spintube_chirality_action}
We now focus on the action $S_{ch}$ for the fields $\phi_1$ and
$\phi_2$, which stands for the chirality (non-magnetic) degree of
freedom we will introduce in Sec~\ref{sec:strongly_coupled_limit}.  Before going into the
technical details, a first comment is in order here. As we mentioned
before, the chosen classical configuration on top of which the path
integral approach is constructed explicitly breaks the $\mathbb{Z}_2$
symmetry related to chirality. However, from the arguments shown below,
we expect it to be able to reproduce almost all phases potentially observable 
for the spin tube.

It is convenient to introduce the two complex conjugate fields $\Psi=\phi_1+i\phi_2$ and 
$\Psi^{*}=\phi_1-i\phi_2$ and, after rescaling the time, we get the action
\begin{equation}
\begin{split}
S[\Psi,\Psi^{*}]=\int d\tau dx \bigg\{ &\frac{1}{2}K|\vec{\nabla}\Psi|^{2} + \tilde{M}^2|\Psi|^2 \bigg.\\
&-2\mu(\Psi^{*}\partial_{\tau}\Psi - \Psi\partial_{\tau}\Psi^{*})+\ldots \bigg\},
\end{split}
\label{eq:simple_spintube_chirality_action}
\end{equation}
where $K=\sqrt{\lambda_{\tau}^{(1,2)}\lambda_{x}^{(1,2)}}$ and  
$\tilde{M}^2=M^2\sqrt{\frac{\lambda_{\tau}^{(1,2)}}{\lambda_{x}^{(1,2)}}}$. We write $(\ldots)$ 
to remind that we are currently working in a second order expansion and higher order terms are 
expected in the general action. Although there is a mass term, we believe that the Berry term 
$\mu(\Psi^{*}\partial_{\tau}\Psi - \Psi\partial_{\tau}\Psi^{*})$ may induce strong effects 
on the behaviour of $\Psi$, namely the possibility to have a gapless phase. 
We propose to treat qualitatively this question by using the symmetries 
to write a general action including important higher order terms.

Going back to the initial fluctuation variables defined in 
(\ref{eq:simple_spintube_flucuations}), it is instructive to rewrite the field $\Psi$ as 
$\Psi=i\frac{2}{\sqrt{6}}(\varphi_3 + \omega\varphi_1 + \omega^2\varphi_2)$, where $\omega=e^{i\frac{2\pi}{3}}$. 
So under a circular permutation of the sites on a triangle, which let the system invariant, 
the field $\Psi$ grabs a phase factor $\Psi \rightarrow \omega\Psi$. Then the most general  
action invariant under such transformations has the form
\begin{equation}
\begin{split}
S[\Psi,\Psi^{*}]=\int d\tau dx &\bigg\{ \frac{1}{2}K|\vec{\nabla}\Psi|^{2} - \mu(\Psi^{*}\partial_{\tau}\Psi - \Psi\partial_{\tau}\Psi^{*}) \bigg.\\ 
&+ \beta\Psi^3+\beta^*\Psi^{*3} + f(|\Psi|,\phi_s) + \ldots \bigg\}.
\end{split}
\label{eq:simple_spintube_chirality_general}
\end{equation}
Writing $\Psi=\rho e^{i\theta}$, the only potentially gapless degree of freedom is 
the phase field $\theta$, and the most general action reads
\begin{equation}
\begin{split}
S[\theta]=\int d\tau dx \bigg\{ &\frac{1}{2}\tilde{K}|\vec{\nabla}\theta|^{2} + \lambda_3\mathrm{cos}(3\theta) + 
\lambda_6\mathrm{cos}(6\theta) \bigg.\\
&+i\mu\partial_{\tau}\theta \bigg\}.
\end{split}
\label{eq:simple_spintube_phase_action}
\end{equation}
where $\tilde{K}$, $\lambda_3$, $\lambda_6$ and $\mu$ are phenomenological parameters. The 
important point to note is that the original Berry phase causes the 
$i\mu\partial_{\tau}\theta$ term that forbid the vorticity (see the discussion for the 
symmetric action). 

The effective action (\ref{eq:simple_spintube_phase_action}) tells us that there 
are four possible phases :

{\it i)} When the stiffness $\tilde{K}$ of the field $\theta$ is large
enough, the scaling dimensions $\Delta_{3,6}\sim 1/\tilde{K}$ of the
cosines are small and they are relevant so that $\langle \Psi
\rangle \neq 0$. We first assume that $\lambda_6$ has the same sign as
$\lambda_3$. In this case we have only three equivalent solutions for
$\langle \Psi \rangle$ in which two of the three fields $\varphi_i$
have the same value.  To understand the consequences of this, let us
go back to eq.  (\ref{eq:simple_spintube_low-energy_action}) and
rewrite the before the last term as
$\Pi_1(\varphi_3-\varphi_2)+\Pi_2(\varphi_1-\varphi_3)+\Pi_3(\varphi_2-\varphi_1)
\propto i(\Pi\Psi^*-\Pi^*\Psi)$, where the complex conjugate fields
$\Pi$ and $\Pi^*$ have the same definition than the fields $\Psi$,
$\Psi^*$ but with respect to the fluctuations $\Pi_{\alpha}$. Having
$\langle \Psi \rangle \neq 0$ implies $\langle \Pi \rangle \neq 0$,
which translates into a homogeneous renormalization of the classical
value of the magnetization.  This correction is equal for two spins
but different for the third one (note that the total magnetization is
kept unchanged). We call this phase the symmetric spin imbalance.

{\it ii)} The cosine operators are again relevant so that $\langle \Psi \rangle \neq 0$ 
but now  $\lambda_6$ has the opposite  sign than $\lambda_3$. As for the phase ({\it i}) 
there is an homogeneous spin imbalance but with three different values for the magnetization, 
and we dub this phase an asymmetric spin imbalance phase. The transition from the threefold degeneracy 
of the phase ({\it i}) to the sixfold degeneracy of this phase corresponds to the double sine-Gordon model,
so it belongs to the universality class of the Ising transition.~\cite{Delfino1998}

{\it iii)} For a sufficiently small stiffness, the cosine operators become irrelevant. 
In that case we perform a duality transformation as previously to take into account the 
role of the i$\partial_{\tau}\theta$ term. We end with two terms 
$(1/\tilde{K})(\vec{\nabla}\Theta)^2$ and $\lambda_2\mathrm{cos}(2\pi(\Theta+\eta x))$, 
where $\Theta$ is the dual field of $\theta$ and $\eta$ a phase modulation, {\it a priori} function 
of the microscopic parameters. For general $\eta$ phases, the cosine is not 
commensurate and that eventually leads to a gaussian model. This phase is characterized by 
$\langle \Psi \rangle = 0$ and algebraic correlation functions for the $\theta$ field. We have a 
conformal field theory with central charge (see Sec.~\ref{sec:simple_spintube_DMRG_entropies}) $c=1$.

{\it iv)} $\langle \Psi \rangle = 0$ and the correlation functions for the $\theta$ 
field are short-ranged. Here some comments are in order concerning the action 
(\ref{eq:simple_spintube_phase_action}). The last term, which originates from the Berry 
phase, has the effect to suppress vortex configurations for the field $\theta$ and is responsible 
for the gapless phase ({\it iii}). It is also the same scenario found to explain 
that the field $\phi_s$ is gapless in general except for particular values of the 
magnetization. In the present case, as long as the parameter $\mu$ has a generic value, 
we said that the cosine operator of the dual field  is forbidden.~\footnote{Note that a fine-tuning 
of the modulation may lead to $\mathbb{Z}_3$ criticality, such as the one
found in spin-1/2 chain in magnetic field, see P. Lecheminant and E. Orignac, Phys. Rev. B {\bf 69}, 174409  (2004).}
 Thus the transition from the phase 
({\it iii}) to this short-ranged phase should not be in principle via a 
Berezinsky-Kosterlitz-Thouless (BKT) transition. It would be rather because $\tilde{K} \rightarrow 0$, much in 
the same manner than a XXZ chain enters into the ferromagnetic phase when the exchange anisotropy parameter becomes sufficiently negative. Such
a ferrochiral phase has been found for $S=1/2$ and weakly coupled chains in a wide range of the magnetic field by 
Sato in Ref.~\onlinecite{Sato2007}. He argued that this order should also survive when entering the plateau state 
for a moderately larger rung coupling.

\subsection{Strongly coupled chains : effective models and chirality}\label{sec:simple_spintube_eff_chir}
\subsubsection{First-order perturbation Hamiltonians}\label{sec:strongly_coupled_limit}
In Sec.~\ref{sec:simple_spintube_path}, we have discussed the model (\ref{eq:simple_spintube_Hamiltonian}) 
regardless of the value of the spin $S$ or the strength of
the coupling parameters $J_{\parallel}$ and $J_{\perp}$. We eventually found 
a condition on the magnetization plateau values. In this section, we
focus on the half-integer spin case and on the strong coupling limit
between the chains, or weakly coupled triangles,
$J_{\perp}/J_{\parallel}\to\infty$. For a given half-integer $S$, we study the lowest and
highest magnetization plateaux, namely $m=1/6$ and $m=S-1/3$, as we will show they both can be described in terms
of an additional chirality degree of freedom.

We start with the $S=1/2$ case in the strong coupling limit, and consider first the extreme 
case $J_{\parallel}=0$ where the system is made of decoupled triangles. The ground-state 
of a triangle is fourfold degenerate at $h=0$, with two chiral spin-$1/2$ doublets. These states are
\begin{equation}
\begin{split}
|\uparrow L \rangle=\frac{1}{\sqrt{3}}\left(|\uparrow \uparrow
\downarrow\rangle +\omega |\uparrow \downarrow \uparrow \rangle+\omega^{-1} |\downarrow \uparrow \uparrow\rangle\right),\\
|\downarrow L \rangle=\frac{1}{\sqrt{3}}\left(|\downarrow \downarrow
\uparrow\rangle +\omega |\downarrow \uparrow \downarrow \rangle+\omega^{-1} |\uparrow \downarrow \downarrow\rangle\right),\\
|\uparrow R \rangle=\frac{1}{\sqrt{3}}\left(|\uparrow \uparrow
\downarrow\rangle +\omega^{-1} |\uparrow \downarrow \uparrow \rangle+\omega |\downarrow \uparrow \uparrow\rangle\right),\\
|\downarrow R \rangle=\frac{1}{\sqrt{3}}\left(|\downarrow \downarrow
\uparrow\rangle +\omega^{-1} |\downarrow \uparrow \downarrow \rangle+\omega |\uparrow \downarrow \downarrow\rangle\right),
\end{split}
\label{eq:ground_state_one_triangle}
\end{equation}
where $\omega=e^{i\frac{2\pi}{3}}$. The indices $L$ and $R$ represent
the chirality and $\uparrow,\downarrow$ the z-axis
projection of the total spin of the triangle. It is important here to make 
the link with the field $\Psi\propto \varphi_3 + \omega\varphi_1 + \omega^2\varphi_2$ 
defined in the path integral approach. Given the similar form of the states 
(\ref{eq:ground_state_one_triangle}), it indicates clearly that the field $\Psi$ 
(or equivalently the fields $\phi_1$ and $\phi_2$) describes this 
chirality degree of freedom. This is also consistent with the fact that the 
Berry phase of $S_{ch}$ in (\ref{eq:simple_spintube_lowenergy_action_pi_integrated}) 
disappears in the opposite limit $J_{\perp}/J_{\parallel}\to 0$.

In this strong coupling limit, we keep only 
the four states (\ref{eq:ground_state_one_triangle}) to describe the low-energy 
physics around the zero magnetic field level crossing. To first-order in
$J_{\parallel}/J_{\perp}$ the effective Hamiltonian reads
\begin{equation}
H_{eff}=\frac{J_{\parallel}}{3}\sum_{j} [1+4 (\tau ^{+}_{j}\tau^{-}_{j+1}+ \tau ^{-}_{j}\tau^{+}_{j+1})]\vec{S}_j.\vec{S}_{j+1} - h\sum_j S^z_j,
\label{eq:simple_spintube_effective_Hamiltonian_S1_2_plateau_1_6}
\end{equation}
where $\vec{S}_j$ is the triangle total spin-$1/2$ operator. We define
the pseudo-spin-1/2 chirality operators $\tau^{\pm}_j$. They
exchange chiralities $L$ and $R$ such as
\begin{equation}
\begin{split}
&\tau^{+}|\cdot L\rangle=0, \quad \quad \quad  \tau^{-}|\cdot
L\rangle=|\cdot R\rangle,\\
&\tau^{+}|\cdot R\rangle=|\cdot L\rangle, \quad\ \tau^{-}|\cdot R\rangle=0 .
\end{split}
\label{eq:chirality_operators}
\end{equation}
By construction, this effective Hamiltonian describes the system from zero
magnetization up to the first plateau $m=1/6$ with $\langle S^z_i \rangle=+1/2$,
where only the two polarized states remain. This model has been studied both
analytically and numerically.~\cite{Schulz1997,Kawano1997,Orignac2000} Its spectrum 
displays a small plateau at magnetization $m=0$, the spin gap arising
from the dimerization of the ground state as explained in Sec.~\ref{sec:intro}. 
A strong enough magnetic field closes the gap and the
system is then described by a two-component Luttinger liquid, with both
spin and chirality modes being gapless.~\cite{Orignac2000} Increasing
again the magnetic field drives the system to the magnetization
plateaux where only the two $S^z = +1/2$ states are present. The chirality
is described by the XY Hamiltonian 
\begin{equation}
H_{eff}=\frac{J_{\parallel}}{12}\sum_{j} [1+4 (\tau ^{+}_{j}\tau^{-}_{j+1}+ \tau ^{-}_{j}\tau^{+}_{j+1})].
\label{eq:spintube_Heff_l2}
\end{equation}
and is then critical.

This description remains valid for higher half-integer spins
$S$. The low-energy space of one triangle at zero magnetic field is always
spanned by two degenerate chiral doublets whose spin projections are
$S^z=\pm 1/2$, so the above description used to derive the effective
Hamiltonian can be repeated. The region from zero field up to the
first plateau $m=1/6$ is described by an Hamiltonian of the form
(\ref{eq:spintube_Heff_l2}), with only a change in the numerical constant
of the chirality operators. On the plateau, the physical spin is frozen to +1/2 and 
the chirality is governed by an XY model which reads
\begin{equation}
H_{eff}=\frac{J_{\parallel}}{12}\sum_{j} [1+\alpha (\tau ^{+}_{j}\tau^{-}_{j+1}+ \tau ^{-}_{j}\tau^{+}_{j+1})],
\label{eq:spintube_Heff_S}
\end{equation}
where the single parameter is $\alpha=(2S+1)^2$.

Starting again from the decoupled case $J_{\parallel}=0$, we observe
that the above chirality description can also be used for the
highest magnetization plateau $m=S-1/3$, where the isolated triangle
ground state is also twofold degenerate. Using again first-order
perturbation theory, we find the chirality states on the plateau are
also given in terms of a XY model of the same form as (\ref{eq:spintube_Heff_S}).

\subsubsection{Range-2 CORE Hamiltonians}\label{sec:simple_spintube_range_2_CORE}
However, it turns out that these first-order effective Hamiltonians do
not capture entirely the behaviour of the chirality on these two
extreme plateaux, as we observed numerically by measuring the central
charges the existence of gapped phases for some range of the coupling (see Sec.~\ref{sec:simple_spintube_DMRG_entropies}).

A way to go beyond the simple first-order perturbation theory is to use a 
Contractor Renormalization~\cite{Morningstar1996} (CORE) 
approach to compute numerically an effective Hamiltonian. The CORE technique is a 
non-perturbative method of renormalization in real space for lattice systems, used to build
effective Hamiltonians reproducing the low-energy physics. It has 
been shown to give quantitative results for instance for various antiferromagnetic models~\cite{Capponi2004}, including the presence of 
 magnetic field.~\cite{Abendschein2007} 
Here, we truncate the calculation of the effective interactions to range-2, i.e. we consider only two coupled triangles. This is a quite simple computation, but it already gives a qualitative improvement over lowest-order perturbation, 
since it produces an effective Hamiltonian of the XXZ type~:
\begin{equation}
H_{eff}=\sum_{j}\bigg[ \frac{J_{xy}}{2}(\tau ^{+}_{j}\tau^{-}_{j+1}+ \tau ^{-}_{j}\tau^{+}_{j+1}) + J_z\tau ^{z}_{j}\tau^{z}_{j+1} \bigg].
\label{eq:simple_spintube_Heff_range}
\end{equation}

\begin{figure}[!ht]
\begin{center}
\includegraphics[width=0.45\textwidth,clip]{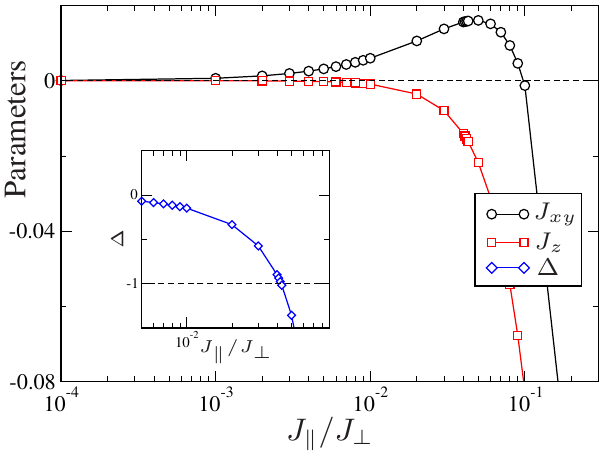}
\end{center}
\caption{(Color online) Values of the effective XXZ Hamiltonian parameters $J_{xy}$, 
$J_z$ and $\Delta=J_z/J_{xy}$ describing the chirality on the 
$m=7/6$ plateau for a spin tube $S=3/2$. The inset shows that at a 
critical coupling $J_{\parallel}/J_{\perp}=0.042$ the chirality 
excitations become gapped and the model enters the ferrochiral phase.}
\label{Fig:simple_spintube_parameters_XXZ_S32}
\end{figure}

We show in Fig.~\ref{Fig:simple_spintube_parameters_XXZ_S32} the
computed values of the parameters $J_{xy}$, $J_z$ and their ratio
$\Delta=J_z/J_{xy}$ governing the behaviour of the model
(\ref{eq:simple_spintube_Heff_range}), for the plateau $m=7/6$ in the
case $S=3/2$. For coupling values $J_{\parallel}/J_{\perp}<0.042$, the
system is in the regime $|\Delta|<1$, the gapless XY phase with a
central charge $c=1$. As the coupling is increased, we see that the
negative $J_z$ component decreases and at the critical value
$J_{\parallel,c}^h/J_{\perp}=0.042$, the system enters in the regime
$\Delta<-1$, corresponding to the ferromagnetic (``ferrochiral'' here)
phase $c=0$ where all the triangles have the same chirality $L$ or
$R$. Note that the change of sign of $J_{xy}$ does not change the
nature of the phase as only the sign of $J_z$, which remains negative,
is important.  Also, we have checked by exact diagonalization (ED) for
larger system lengths (up to $L=10$) that there is a level crossing
close to this critical coupling. The quantum numbers of the ground state 
are compatible with the XY to ferrochiral scenario. Finally, the same scenario for
$\Delta$ occurs for the lowest plateau $m=1/6$ at a critical coupling
$J_{\parallel,c}^l/J_{\perp}=0.256$ (however we will see with the DMRG results 
that for this value we are no longer in the plateau phase).

Even if this approach is straightforward, it allows to explain qualitatively  
 the possibility of a phase transition from a critical to a gapped phase.
On the quantitative side, although longer range effective interactions are expected to play a role (see below),  the critical value 
$J_{\parallel}/J_{\perp}<0.042$ found here is 
very close to the transition value observed with 
the DRMG.

As this non-perturbative CORE computation only involves solving two triangles, 
we can also treat higher spins. The Table~\ref{Tab:simple_spintube_chirality_critical_values} 
shows, for different values of the half-integer spin $S$, the
critical values $J_{\parallel,c}^l$ and $J_{\parallel,c}^h$ at which
the chirality on the lowest and the highest plateaux is expected to 
undergo a quantum phase transition from the XY phase to the Ising phase. 
For both plateaux, we observe the gapless phase shrinks in the
large $S$ limit. Again, the results given by this simple method are encouraging 
and we believe them to be actually quite accurate. Indeed, in the
spin-$1/2$ case we find the value 0.500, very close to the
value 0.496 computed with DMRG where the chirality enters the ferrochiral phase~\cite{Okunishi2012}. 
In the following we will present DMRG data supporting these results for $S=3/2$.
\begin{table}[htbp]
\begin{center}
\begin{tabular}{|c|c|c|c|c|c|}
\hline  & $S=1/2$ & $S=3/2$ & $S=5/2$ & $S=7/2$ & $S=9/2$ \\ 
\hline $J_{\parallel,c}^l$ & 0.500 & 0.256 & 0.157 & 0.108 & 0.079 \\ 
\hline $J_{\parallel,c}^h$ & 0.500 & 0.0420 & 0.0140 & 0.0071 & 0.0042 \\ 
\hline 
\end{tabular}
\caption{Critical values $J_{\parallel,c}^l$ and $J_{\parallel,c}^h$ of 
the transition from an XY effective model to a ferromagnetic Ising one 
to describe the chirality behaivour on the lowest and the highest 
magnetization plateaux.}
\label{Tab:simple_spintube_chirality_critical_values}
\end{center}
\end{table}

\subsubsection{General effective Hamiltonians}\label{sec:general_effective_Hamiltonians}
Recently, Okunishi {\it et al.} have derived the second-order effective Hamiltonian in
the spin-1/2 case~\cite{Okunishi2012}, which is a special case since 
it has only one plateau. New terms appear, such as a negative $\tau^z\tau^z$ one 
which can drive the system into an ordered phase. This is in agreement with the CORE calculation.
Our goal is to propose an effective Hamiltonian capturing these phases and the
transition. As the argument of the phase factor $\omega$ in 
Eq.~(\ref{eq:ground_state_one_triangle}) is nothing else than the transerve momentum, the
effect of the operators $\tau^{\pm}$ is simply to shift the triangle
momentum of $\pm 2\pi/3$. By transverse momentum conservation, 
the most general effective Hamiltonian, on both the first and 
last plateau, can be written as 
\begin{equation}
\begin{split}
H_{eff}=\sum_{j}\bigg[ &\frac{J_{xy}}{2}(\tau ^{+}_{j}\tau^{-}_{j+1}+ \tau ^{-}_{j}\tau^{+}_{j+1}) + J_z\tau ^{z}_{j}\tau^{z}_{j+1} \bigg.\\
&+ J_3(\tau^{+}_{j-1}\tau ^{+}_{j}\tau^{+}_{j+1} + \tau^{-}_{j-1}\tau ^{-}_{j}\tau^{-}_{j+1}) + \ldots \bigg],
\end{split}
\label{eq:simple_spintube_Heff_generic}
\end{equation}
where we have dropped, for example, second-neighbour exchange terms. The values 
of the parameters of this model have been calculated up to second-order in $J_{\parallel}/J_{\perp}$ (see~\onlinecite{Okunishi2012}). 
The $\tau ^{+}\tau ^{+}\tau ^{+}$ term was obviously absent in our range-2 CORE calculation
but we would except it to appear for a higher range one. 
Notice that in this language the $\mathbb{Z}_2$ symmetry associated with chirality 
is just $\tau^{z} \to - \tau^z$ which can be obtained for example with a 
rotation of $\pi$ around the $x$ axis. At this point one can try to make 
connection with the results of the path integral. We have to keep in mind 
that, by construction, the  $\mathbb{Z}_2$ chirality symmetry is broken within 
the path integral approach. This corresponds to placing the effective spin chain 
above at a non-zero average homogeneous magnetization $\langle \tau^{z} \rangle \neq 0$. 

The bosonized form of this effective Hamiltonian was written recently in Ref.~\onlinecite{Okunishi2012} as
\begin{eqnarray}
H = { v \over 2} \int dx ~ \left[ {1\over 2 \kappa } (\partial_x \chi)^2 + 2 \kappa  (\partial_x \tilde{\chi})^2 \right. \nonumber \\
\left.+ \lambda_1 \cos(2 \sqrt{2\pi } \chi) + \lambda_2 \cos(6 \sqrt{2\pi } \tilde{\chi}) \right],
\label{eq:simple_spintube_bosonized_effective_Hamiltonian}
\end{eqnarray}
where $v$ is a Fermi velocity, $\kappa$ is the Luttinger parameter and
$\tilde{\chi}$ is dual to ${\chi}$. The second cosine operator is
radiatively generated by the presence of the $\tau^+\tau^+\tau^+ +
~h.c. $ term. The first cosine operator is relevant for $\kappa < 1$
while the second becomes relevant for $\kappa > 9$.  The gapless phase
obtained between these two critical points is associated with the case
({\it iii}) predicted by the path integral in Sec.~\ref{sec:simple_spintube_chirality_action}.  For
$J_{\parallel}/J_{\perp}\rightarrow0$, the effective Hamiltonian
(\ref{eq:simple_spintube_Heff_generic}) reduces to an XY model
corresponding to $\kappa =2$, thus both cosine terms are irrelevant
and the chirality is in a critical phase. When the coupling ratio
$J_{\parallel}/J_{\perp}$ is increased, it turns out that $\kappa$
increases too as we are in the ferromagnetic regime, so the first
cosine $\cos(2 \sqrt{2\pi } \chi)$ will always be irrelevant.~\footnote{In the case where both cosines are in 
competition, an exotic criticality can emerge, see for example P. Lecheminant, A. O. Gogolin and A. A. Nersesyan, Nucl. Phys. B {\bf 639}, 502 (2002).} We
predict the gapless phase to disappear and two gapped phases should
appear successively. The first one is caused by the $\cos(6
\sqrt{2\pi } \tilde{\chi})$ term becoming relevant for some critical
negative $\Delta_c$, and the second corresponds to a transition
to the ferrochiral phase when the magnitude of the negative $\Delta$
becomes sufficiently large.  We associate this second phase to the
case ({\it iv}) predicted by the path integral approach. The first case
eventually leads to the appearance of a gap in the chirality degrees of
freedom in favor of a spin imbalance phase similar to the one found
in Ref.~\onlinecite{Okunishi2012} but with one-step breaking of the translation
symmetry.

Indeed, having a non-zero expectation value for the field $\tilde{\chi})$ make that the 
operator $\tau^x\sim (-1)^x\mathrm{cos}(\tilde{\chi})$ is also non-zero. This correponds 
to a symmetry breaking, as this operator is directly 
related to an imbalance in terms of the original spin operators.~\footnote{In this staggered spin-imbalance phase, two sites per rung have the same magnetization so that 
reflection symmetry is preserved and the ground-state is 6-fold degenerate. Note that the presence of a higher harmonic $\cos (12 \sqrt{2\pi } \tilde{\chi})$ may 
induce a fully symmetry-broken spin-imbalance phase with all three magnetizations different (i.e. 12-fold degenerate groundstate).
} As this phase 
is incompatible with a non-zero average magnetization in the $z$ direction, 
it is inaccessible to the path integral approach we have presented before.
Also, the critical point at which the spin 
imbalance phase would occur is expected to be very close to the regime in which the ferrochiral 
phase appears (which, in the absence of the three-body $\tau^+\tau^+\tau^+ + ~h.c. $ term, 
occurs for $\Delta=-1$). We the believe this phase to be 
only present in a very narrow range between the XY and the ferrochiral phases.

\subsection{DMRG results for $S=3/2$}\label{sec:DMRG_simple_spintube}
\subsubsection{Magnetization plateaux}\label{sec:simple_tube_DMRG_plateaux}
In order to verify the previous predictions about the magnetization
plateaux, which were established thanks to large-$S$ techniques, we
have performed numerical simulations using the DMRG algorithm~\cite{White1992}
for $S=3/2$ spin tubes with open boundary conditions (OBC) along the
legs. We consider system lengths up to $L=64$. Typically, we kept up
to 2000 states and perform 20 sweeps, 
which is sufficient to have a discarded weight smaller
than $10^{-8}$ or less.

\begin{figure}[!htb]
\includegraphics[width=0.45\textwidth,clip]{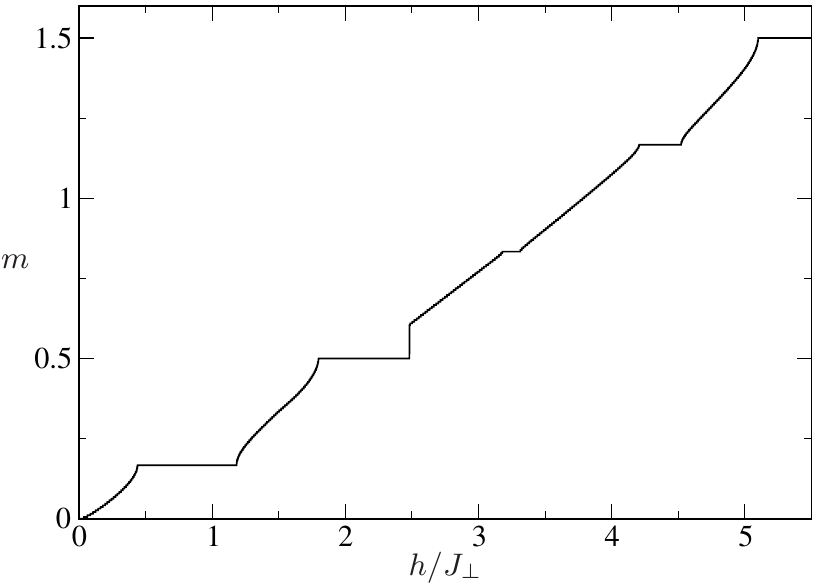}
\caption{Magnetization per spin curve ($m$) vs magnetic field 
$h$ for the simple spin tube in the case $S=3/2$. DMRG simulations 
were performed with $L=64$ and $J_\parallel/J_\perp=0.1$. Finite-size steps 
are almost not visible on this scale.}
\label{fig:magnetization_simple_spintube_S32}
\end{figure}

In Fig.~\ref{fig:magnetization_simple_spintube_S32}, we plot a typical magnetization
curve obtained in the strongly-coupled chains regime for $J_\parallel/J_\perp=0.1$. 
Large plateaux are observed below saturation for magnetization per site $m=1/6$, $3/6$, $5/6$ and
$7/6$, which correspond to the condition (\ref{eq:simple_spintube_plateau_condition}) 
that we have found with the field theory. Remember
also that an $m=0$ plateau was predicted, but it has a different nature
(dimerization of the ground-state), and on the scale of the figure, it
is hardly visible.

\begin{figure}[t!]
\includegraphics[width=\columnwidth,clip]{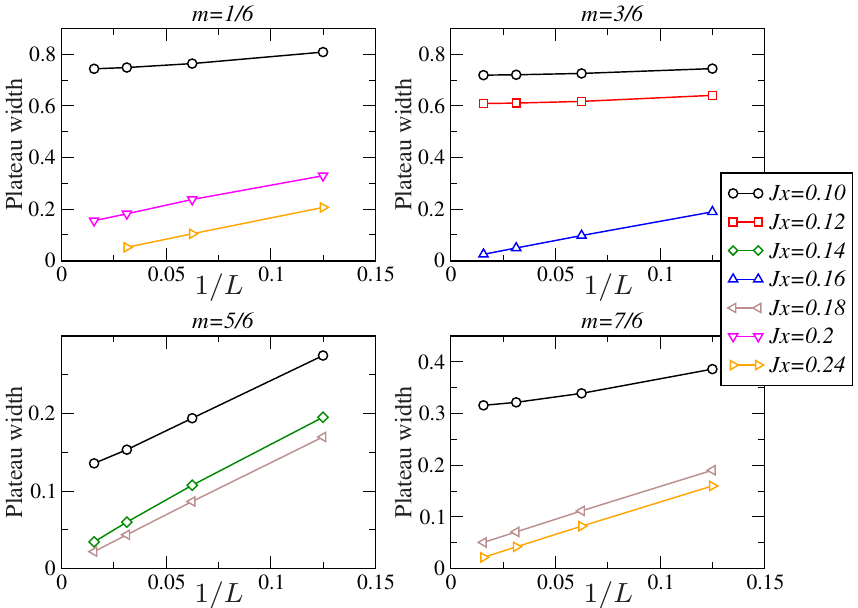}
\caption{(Color online) Finite-size scaling of the plateaux widths for different 
couplings $J_\parallel/J_\perp$.}
\label{fig:SizePlateaux}
\end{figure}

\begin{figure}[t!]
\includegraphics[width=0.45\textwidth,clip]{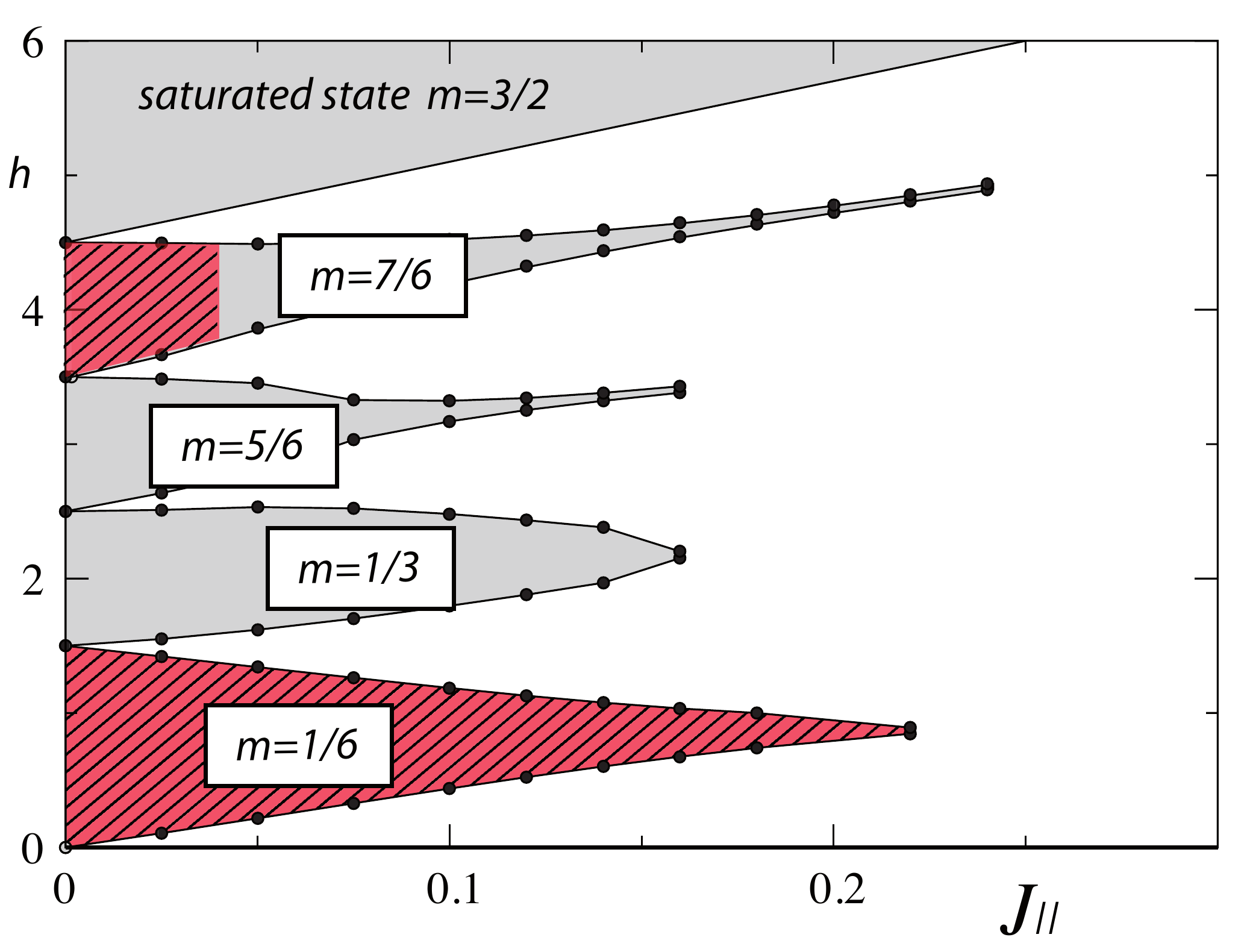}
\caption{(Color online) Phase diagram of the three-leg spin tube with $S=3/2$ 
as a function of the coupling $J_\parallel$ and magnetic field $h$ ($J_{\perp}=1$ 
is the unit of energy). Several magnetization plateaux can be observed (filled areas) 
and an additional $m=0$ plateau is found (bold line). Data correspond to numerical 
simulations on $3\times 32$ lattice with DMRG. Inside the extreme plateaux 
$m=1/6$ and $m=7/6$, hashed red regions correspond to critical chirality phases 
(see Sec.~\ref{sec:simple_spintube_DMRG_entropies}).}
\label{fig:simple_spintube_phase_diag_S32}
\end{figure}

In order to map out the phase diagram, we perform a finite-size
analysis of the widths of each plateau for several
couplings. Resulting data are shown in
Fig.~\ref{fig:SizePlateaux}. While it confirms that all the plateaux
found for $J_\parallel/J_\perp=0.1$ are present in the thermodynamic
limit, we do observe that each of them disappears for some different critical
ratios of the coupling constants. These critical values of $J_{\parallel}/J_{\perp}$ 
are summarized in the Table~\ref{tab:simple_spintube_plateaux_scaling} along 
with the predicted values coming from the formula (\ref{eq:simple_spintube_cosine_scalingdim}). 
It is important to mention that because the transitions between 
the plateau phases and the gapless phases are expected to be of the 
BKT type, it is difficult to locate 
them accurately. We also note that the predicted values are in a roughly 
good agreement with those coming from the DMRG. The main qualitative difference 
is that the path integral approach predicts that the highest plateau, $m=7/6$ here, 
should be the most robust. This prediction of the plateaux disappearing 
gradually with $m$ seems to be a general feature of the path 
integral approach for chains or coupled chains. 
By performing extensive simulations, we arrive at the phase 
diagram shown in Fig.~\ref{fig:simple_spintube_phase_diag_S32}, 
where we indicate both the plateaux phases and the chirality phases 
discussed in the next subsection. It is clear that the predicted 
disappearance pattern for the plateaux is not recovered exactly, 
but we have to remember that the result (\ref{eq:simple_spintube_cosine_scalingdim}) 
comes from a large-$S$ approach.
 We 
plot the saturation field $h_{sat}$, which can be found  analytically to be 
$h_{sat}=(3J_\perp +4 J_\parallel)S$. We also indicate 
the existence of the $m=0$ plateau, which we observe on the whole range 
of $J_{\parallel}/J_{\perp}$ but is not visible on this scale. 
For instance, we find the spin gap to be of order $2.10^{-2}J_{\perp}$ for 
$J_{\parallel}/J_{\perp}=0.1$, a value in good agreement with other DMRG 
calculations.~\cite{Nishimoto2011}

\begin{table}[htbp]
\begin{center}
\begin{tabular}{|c|c|c|c|c|}
\hline  & $m=1/6$ & $m=3/6$ & $m=5/6$ & $m=7/6$ \\ 
\hline $J_{\parallel,c}^{DMRG}$ & 0.20-0.22 & 0.14-0.16 & 0.16-0.18 & 0.22-0.26 \\ 
\hline $J_{\parallel,c}^{PI}$ & 0,066 & 0.074 & 0.098 & 0.19 \\ 
\hline 
\end{tabular}
\caption{Critical values $J_{\parallel,c}$ for the plateaux, in unit of 
$J_{\perp}$, for the spin tube with $S=3/2$. We indicate both the values found 
by DMRG and with the path integral.}
\label{tab:simple_spintube_plateaux_scaling}
\end{center}
\end{table}

\subsubsection{Entanglement entropies and conformal charges}\label{sec:simple_spintube_DMRG_entropies}
Since the chirality degree of freedom has been shown to emerge for the
extreme plateaux for half-integer spin, we now use large-scale DMRG simulations to investigate it. 
For $S=1/2$, chirality is only expected on $m=1/3$ plateau, and has already been
confirmed numerically.~\cite{Okunishi2012} In our paper, we
consider the next case, i.e. $S=3/2$.

In order to check the existence of a chirality phase transition on the
extreme plateaux, we simply compute the block von Neumann entropy
$S_{vN}(\ell)$ which exhibit two different behaviours for large blocks
$\ell$ and OBC: $S_{vN}(\ell)$ saturates to a constant when the system
is fully gapped, whereas $S_{vN}(\ell) \simeq (c/6) \log \ell$ where $c$
is the central charge of the underlying conformal field
theory.~\cite{Calabrese2004} In order to minimize finite-size effects, 
we will consider the conformal block length $d(\ell|L)=(L/\pi) \sin (\ell\pi/L)$. 

\begin{figure}[t!]
\includegraphics[width=0.45\textwidth,clip]{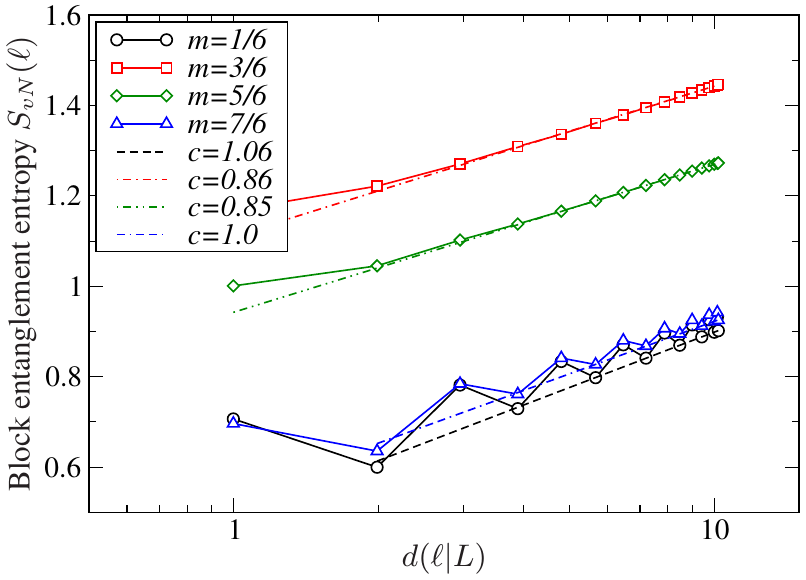}
\caption{(Color online) Block entropy $S_{vN}(\ell)$ vs block length 
$d(\ell|L)$ (starting at one end of the tube) for several magnetization 
plateaux on $L=32$ spin tube. Coupling constants are fixed to 
$J_\parallel/J_\perp=0.02$.}
\label{fig:entropy_Jx_0_02}
\end{figure}

Guided by our CORE analysis, we first choose a small coupling
$J_\parallel/J_\perp=0.02$ where chirality is expected to be gapless
on both $m=1/6$ and $m=7/6$ plateaux. As it is shown in
Fig.~\ref{fig:entropy_Jx_0_02}(a), numerical data are compatible with
a gapless behaviour with $c=1$ in agreement with our expectation.  We
note that the intermediate plateaux $m=3/6$ and $m=5/6$ also possess
critical degrees of freedom, which could be compatible with $c=1$ or a
slightly smaller value. We plan to investigate in the future whether
there could exist a non-gaussian criticality nearby, or if it is
simply due to numerical uncertainty when coupling constants have very
different amplitudes. Anyhow, for intermediate plateaux, there is no
simple chirality language since more than two states per triangle
(respectively four and three) are necessary to describe the low-energy
configurations.

\begin{figure}[!htb]
\includegraphics[width=0.45\textwidth,clip]{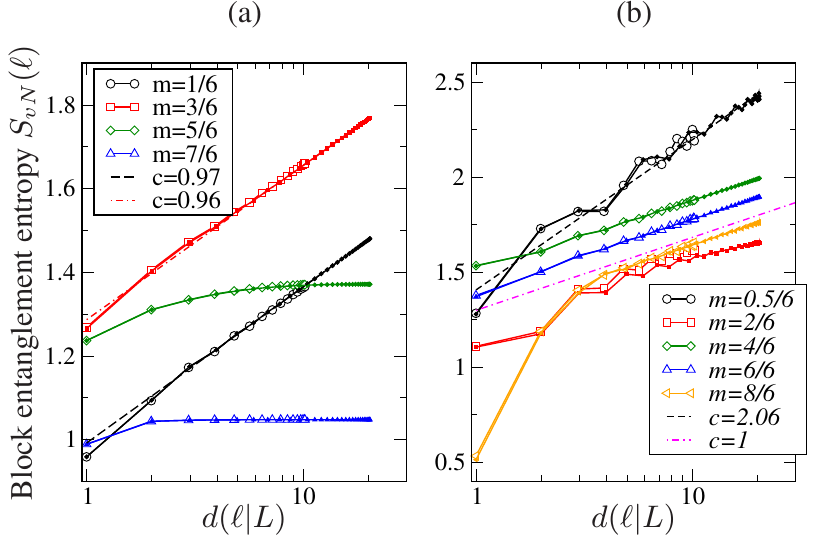}
\caption{(Color online) Block entropy $S_{vN}(\ell)$ vs block length $d(\ell|L)$ 
(starting at one end of the tube) for $L=16$ (open symbols) 
and $L=32$ (filled symbols). Coupling constants are fixed to 
$J_\parallel/J_\perp=0.1$. (a) Magnetizations corresponding to 
plateaux; (b) intermediate magnetizations.}
\label{fig:entropy_Jx_0_1}
\end{figure}

Now, we can increase the coupling constant $J_\parallel/J_\perp$ since
the effective Hamiltonian and the CORE analysis indicate that the chirality degree of freedom should
become gapped beyond some critical values (see Table~\ref{Tab:simple_spintube_chirality_critical_values} 
for $S=3/2$). For instance, when fixing $J_\parallel/J_\perp=0.1$, our data shown in
Fig.~\ref{fig:entropy_Jx_0_1} confirm that chirality has
become gapped in the upper plateau, but remains gapless (with
$c=0.97$) for $m=1/6$.  The critical ratio that we find for gapless
chirality on $m=7/6$ plateau is close to 0.04, while chirality is
always gapless on the $m=1/6$ as long as it exists, i.e.  for
$J_\parallel/J_\perp \lesssim 0.25$. The quantum
phase transition between gapless and gapped chirality phases for the upper 
plateau is indicated on the phase diagram
(Figure~\ref{fig:simple_spintube_phase_diag_S32}).  The central result
is that both critical values are in excellent agreement with our
range-2 CORE estimates (see Table~\ref{Tab:simple_spintube_chirality_critical_values}). 

For completeness, we also plot in Fig.~\ref{fig:entropy_Jx_0_1}(b)
the scaling of the block entropy for intermediate magnetizations, which all correspond of course to critical gapless phases. In particular, at
low magnetization, data are compatible with a 2-component Luttinger
liquid with $c=2$ as predicted.~\cite{Orignac2000} For all the other
magnetizations, our data are compatible with a single gapless mode
$c=1$. It is beyond the scope of this work to study the interplay between chirality and magnetic 
degrees of freedom outside magnetization plateaux, but it could be 
interesting to investigate the stability of the ferrochiral phase for arbitrary magnetic 
field.~\footnote{For the spin 1/2 tube, a ferrochiral phase has been found 
in the weak-coupling regime for any finite magnetic field~\cite{Sato2007}} 

\subsubsection{Nature of the gapped phase}\label{sec:simple_spintube_gapped_phase}
\begin{figure}[!htb]
\includegraphics[width=0.45\textwidth,clip]{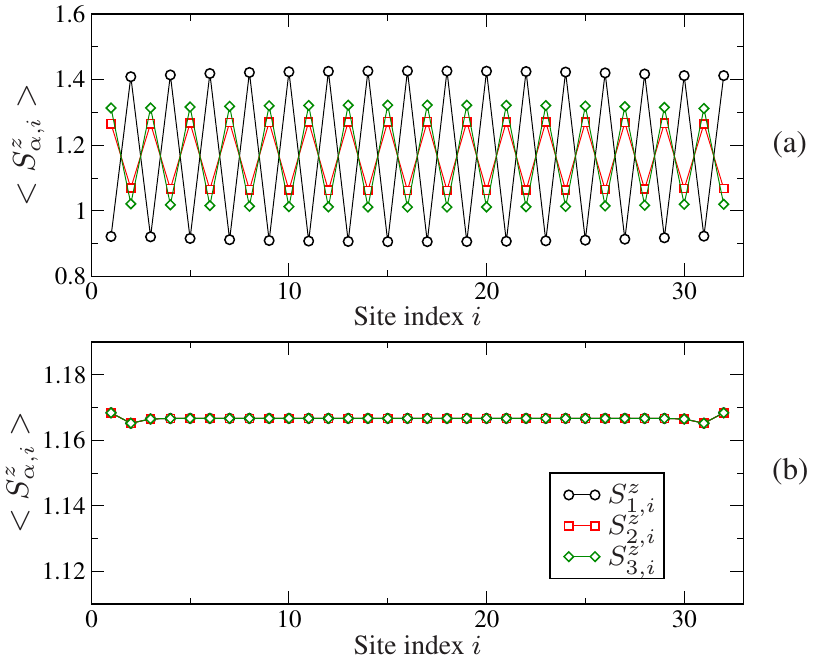}
\caption{(Color online) Local magnetizations obtained by DMRG simulations on a $3\times 32$ spin tube with spin 3/2 on the $m=7/6$ plateau. Upper and lower panels correspond respectively to (a) $J_\parallel/J_\perp=0.04$ and (b) $J_\parallel/J_\perp=0.06$.}
\label{fig:spin_imb}
\end{figure}

From the path-integral approach and the bosonization of the effective
Hamiltonian, several different gapped phases are predicted to possibly
occur when varying the coupling. However, the block entanglement
entropies does not give any information about the nature of the gapped
phase observed for $J_\parallel/J_\perp > 0.04$ in the upper $m=7/6$ plateau. 
To investigate more precisely this question, we have computed the local magnetization
values for different couplings, shown in Fig.~\ref{fig:spin_imb}.

In a very narrow range close to the phase transition, a clear
\emph{staggered} spin imbalance is observed in our simulations (see
Fig.~\ref{fig:spin_imb}a), as predicted from the bosonized Hamiltonian
(\ref{eq:simple_spintube_bosonized_effective_Hamiltonian}). The local
magnetizations vary around their mean value $m=7/6$ with one chain
having a clear different magnetization than the two others (in fact,
we cannot exclude the possibility that all three chains will have
different magnetizations). While the symmetry cannot be broken on a
finite lattice, it turns out that DMRG simulations get locked in one
of the degenerate ground-state. Note also that the level crossing found by ED in this region 
could impede the accuracy of DMRG results here. 
It is interesting to contrast our result with the
small \emph{uniform} spin imbalance phase found in the $S=1/2$
tube which seems to signal the entrance into a regime where the pseudo-spin $1/2$
effective Hamiltonian is not valid anymore.~\cite{Okunishi2012}

When increasing slightly
$J_\parallel/J_\perp$, but still deep in the plateau phase,
we observe that all chains recover the same magnetization (see
Fig.~\ref{fig:spin_imb}b), in agreement with having a ferrochiral
phase.

In order to ascertain the validity of the description in terms of chiral degree of freedom, as described in 
Sec.~\ref{sec:simple_spintube_eff_chir}, we have computed by ED on a small $3\times 6$ cluster with PBC 
the weights of these degrees of freedom in the reduced density matrix of one triangle (see Refs.~\onlinecite{Capponi2004,Abendschein2007} for a 
discussion of this technique). In a wide range  of $J_\parallel/J_\perp \leq 0.08$
including the three different phases, XY, staggered spin-imbalance and
ferrochiral, we find that the weights of these two states exceed $90\%$, so that 
we are rather confident that the effective model in terms of chirality remains valid.


\section{The twisted spin tube}\label{sec:twisted_spintube}
\subsection{The model and its experimental realization}\label{sec:twisted_spintube_model}
\begin{figure}[!htb]
\includegraphics[width=0.26\textwidth,clip]{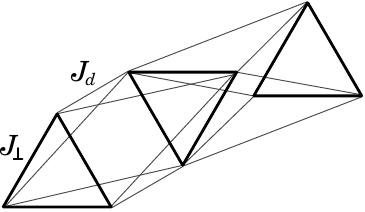}
\caption{Lattice structure of the twisted spin tube.}
\label{fig:twisted_spintube_picture}
\end{figure}

Experimentally, only a few materials have been suggested to realize
spin tube geometries. One such geometry, which we will study in this
section, corresponds to the compound $\mathrm{[(CuCl_2tachH)_3Cl]Cl_2}$.  Magnetic
measurements~\cite{Schnack2004} have shown that it forms a twisted
triangular spin tube. We call it ``twisted'' because of the different 
structure compared to the simple tube of the
Sec.~\ref{sec:simple_spintube}. The spins $S=1/2$, coming from the
copper ions are arranged in a one-dimensional array of equilateral triangles, and
each spin of a triangle is coupled to the spins of the two others chains of the
neighbouring triangles (Fig.~\ref{fig:twisted_spintube_picture}). This 
corresponds to add diagonal couplings to the model 
(\ref{eq:simple_spintube_Hamiltonian}) while the longitudinal one $J_{\parallel}$ 
vanishes. The Hamiltonian describing the twisted spin tube reads
\begin{equation}
\begin{split}
&H=H_{\perp}+H_d+H_h\\
&H_{\perp}=J_{\perp}\sum_j\sum_{\alpha=1,2,3}\vec{S}_{\alpha,j}.\vec{S}_{\alpha+1,j}\\
&H_d=J_{d}\sum_j\sum_{\alpha=1,2,3}\vec{S}_{\alpha,j}.(\vec{S}_{\alpha +1,j+1}+\vec{S}_{\alpha -1,j+1})\\
&H_h=-h\sum_j\sum_{\alpha=1,2,3}S_{\alpha,j}^z,
\end{split}
\label{eq:frustrated_spintube_Hamiltonian}
\end{equation}
where $J_d$ is the diagonal antiferromagnetic coupling. This model is
believed to describe $\mathrm{[(CuCl_2tachH)_3Cl]Cl_2}$ for the values
$J_{\perp}=0.9K$ and $J_d=1.95K$~\cite{Schnack2004}, thus the compound belongs neither
the strong or weak coupling regime. Theorical and experimental 
investigations~\cite{Fouet2006,Ivanov2010} have shown that this compound 
behaves essentially as a spin-3/2 chain, displaying a Luttinger liquid behavior.

\subsection{Path integral approach}\label{sec:twisted_spintube_path}
Following the same steps as in Sec.~\ref{sec:simple_spintube_path}, we start by 
finding the classical ground state of the Hamiltonian (\ref{eq:frustrated_spintube_Hamiltonian}). 
Classically, the extreme cases $J_{\perp}=0$ and $J_d=0$ are easily understood 
in the absence of the magnetic field. For $J_{\perp}=0$, 
the lattice is bipartite and the ground state is the N\'eel state. 
On the other hand, for $J_d=0$ the triangles are decoupled
and, as for the regular spin tube, the three spins are coplanar with a
$2\pi/3$ angle between them. It turns out that the lowest-energy 
configuration is one of those two states in the whole range of the coupling 
parameters.~\cite{Schnack2004} From $J_d/J_{\perp}=0$ to $J_d/J_{\perp}=3/2$, 
the ground-state is the $2\pi/3$ state and for higher
values of $J_d/J_{\perp}$ it changes to the N\'eel state. This can be
seen from a calculation in Fourier space by minimizing the
resulting exchange coupling
$J(k_{\perp},k_{\parallel})=2\mathrm{cos}(k_{\perp})(J_{\perp}+2J_d\mathrm{cos}(k_{\parallel}))$,
where $k_{\perp}=0,2\pi/3$. We have checked numerically that the
effect of a magnetic field is just to polarize the spins in these
two configurations and does not change the transition value
$J_d/J_{\perp}=3/2$. Thus, to investigate this model using the path integral
approach, we have to treat the two regimes separately.

\subsubsection{Regime $J_d/J_{\perp}<3/2$}\label{sec:twisted_spintube_weakly_coupled}
In this range of $J_d/J_{\perp}$, all the triangles are in the same
state, with angles of $120^\circ$ between neighbouring spins partially polarized by 
the magnetic field. This comes from the
three-colorability of the lattice. This state is actually the
same than the umbrella structure for the simple spin tube, but without
the staggered order along the tube. We write this state as
\begin{equation}
\vec{S}_{\alpha,j}=S
\begin{pmatrix} \mathrm{sin}(\theta_0)\mathrm{cos}(\varphi_{\alpha}^0)  \\ \mathrm{sin}(\theta_0)\mathrm{sin}(\varphi_{\alpha}^0) \\ \mathrm{cos}(\theta_0)
\end{pmatrix}
,
\label{eq:frustrated_spintube_classical_gs_inf1.5}
\end{equation}
where $\mathrm{cos}(\theta_0)=\frac{h}{S(3J_{\perp}+6J_{d})}$ and
$\varphi_{\alpha}^0=(\alpha-1)2\pi/3$ up to an additional
constant. Introducing the quantum fluctuations $\theta_0 \rightarrow
\theta_{\alpha,j}=\theta_0 + \delta \theta_{\alpha,j}$,
$\varphi_{\alpha}^0 \rightarrow \varphi_{\alpha}^0+\varphi_{\alpha,j}$
and the conjugate momentum $\Pi_{\alpha,j}$, we get the same expansion 
(\ref{eq:simple_spintube_spin_operators_pi}) for the spin operators, 
except for the alternate order term $(-1)^j$. Then the 
action in the continuum limit reads
\begin{widetext}
\begin{equation}
\begin{split}
S[\{\Pi_{\alpha}\},\{\varphi_{\alpha}\}]=\int d\tau dx \bigg\{&\sum_{\alpha=1,2,3}\left[\frac{1}{2}aJ_{d}(S^2-m^2)(\partial_x\varphi_{\alpha})^2 + \frac{1}{2}a\frac{S^2}{S^2-m^2}(J_{\perp}+2J_{d})\Pi_{\alpha}^2 \right] \bigg.\\
&+a\left(1-\frac{1}{2}\frac{m^2}{S^2-m^2}\right)(J_{\perp}+2J_{d})(\Pi_1\Pi_2+\Pi_2\Pi_3+\Pi_3\Pi_1)\\
&+\frac{1}{4}\frac{S^2-m^2}{a}(J_{\perp}+2J_{d})\left[(\varphi_1-\varphi_2)^2+(\varphi_2-\varphi_3)^2+(\varphi_3-\varphi_1)^2\right]\\
&-\frac{\sqrt{3}}{2}m(J_{\perp}+2J_{d})\left[\Pi_1(\varphi_3-\varphi_2)+\Pi_2(\varphi_1-\varphi_3)+\Pi_3(\varphi_2-\varphi_1)\right]\\
&\bigg. +i\sum_{\alpha=1,2,3}\left[\left(\frac{S-m}{a}\right)\partial_{\tau}\varphi_{\alpha}-\Pi_{\alpha}\partial_{\tau}\varphi_{\alpha}\right]
\bigg\},
\end{split}
\label{eq:twisted_spintube_low-energy_action_inf1.5}
\end{equation}
\end{widetext}

This action has the same form than (\ref{eq:simple_spintube_low-energy_action})
for the simple tube, the only difference appearing in boundary terms. 
We observe that, except for the longitudinal part in $(\partial_x\varphi_{\alpha})^2$, the
couplings $J_{\perp}$ and $J_d$ play the same role. More precisely, at
large-scale, one can argue that the diagonal coupling $J_d$ is essentially identical to 
the perpendicular one. The factor of two for $J_d$ simply tells that there is twice as
many diagonal couplings than perpendicular per unit cell. Thus we can
follow the same steps and after performing the gaussian integration we
find
\begin{widetext}
\begin{equation}
\begin{split}
S[\{\phi_{\alpha}\}]&=S_{ch}[\phi_1,\phi_2]+S_s[\phi_s]\\
S_{ch}[\phi_1,\phi_2]&=\int d\tau dx \bigg\{ \frac{1}{2}\lambda_{\tau}^{(1,2)}\left[(\partial_{\tau}\phi_1)^2+(\partial_{\tau}\phi_2)^2\right]+ \frac{1}{2}\lambda_x^{(1,2)}\left[(\partial_x\phi_1)^2+(\partial_x\phi_2)^2\right] -i\mu(\phi_1\partial_{\tau}\phi_2-\phi_2\partial_{\tau}\phi_1)\bigg\}\\
S_s[\phi_s]&=\int d\tau dx \bigg\{ \frac{1}{2}\lambda_{\tau}^{(s)}(\partial_{\tau}\phi_s)^2+\frac{1}{2}\lambda_{x}^{(s)}(\partial_{x}\phi_s)^2+i3\frac{S-m}{a}\partial_{\tau}\phi_s \bigg\},
\end{split}
\label{eq:twisted_spintube_action_pi_integrated_inf1.5}
\end{equation}
\end{widetext}
where the constants are functions of the microscopic parameters.

At this order the action is again decoupled into a symmetric action 
for $\phi_s$ and a chirality action for the two other
fields $\phi_1$ and $\phi_2$. Surprisingly, the mass of the latters 
vanishes after momentum $\Pi_{\alpha}$ integration. However, it 
does not invalidate our previous discussion about the necessary rescaling of 
$\phi_s$. Indeed, the expansion (\ref{eq:simple_spintube_spin_operators_pi}) 
for the spin operators and for the Hamiltonian is restricted to second 
order in the fields. Expanding up to fourth order, mass terms in
$(\varphi_{\alpha}-\varphi_{\alpha+1})^4$ would appear and we
would recover a mass term, ensuring that $\phi_1$ and $\phi_2$ 
are still small. Appart from this mass cancellation, 
there is no difference with the simple spin tube case. This is expected
as they both have the same classical configuration in this regime of 
$J_d/J_{\perp}$. For the twisted tube, the staggered order 
is in the diagonal coupling, as the angle between spins 
coupled by $J_d$ is larger than $\pi/2$.

Then we perform the duality transformation on the 
symmetric part of the action. We obtain the
plateaux existence condition $3(S-m)\in\mathbb{Z}$, which is obviously
the same as for the simple tube since it does not depend on the detailed geometry 
but only on the unit cell. We also compute the
expression of the cosine operator dimension
\begin{equation}
\Delta=\pi\sqrt{\frac{3J_{d}(S^2-m^2)}{J_{\perp}\left(1+2\frac{J_{d}}{J_{\perp}}\right)}},
\label{eq:frustrated_spintube_cosine_scalingdim_inf1.5}
\end{equation}
which is very similar to the simple tube case (\ref{eq:simple_spintube_cosine_scalingdim}) and 
bears the same functional form with $m$. From the dependence on the microscopic 
parameters, we predict again to observe plateau in the strong coupling 
regime along the rungs.

Concerning the fields $\phi_1$ and $\phi_2$, they describe the chirality degree 
of freedom as for the simple spin tube. The form of the action being the same, 
the same reasoning than in Sec.~\ref{sec:simple_spintube_chirality_action} 
holds and the same four phases are possible. Particularly, we claim again that the Berry 
term in this action makes possible a chiral gapless phase for some values of the microscopic parameters 
we are not able to compute (but still {\it a priori} in the strong coupling regime). Moreover, 
we expect that in the present case, there are more chances to be in this phase compared to 
the simple tube case because of the vanishing bare mass.

\subsubsection{Regime $J_d/J_{\perp}>3/2$}\label{sec:twisted_spintube_strongly_coupled}
We start here from the partially polarized N\'eel state, parametrized as
\begin{equation}
\vec{S}_{\alpha,j}=S
\begin{pmatrix} (-1)^j \mathrm{sin}(\theta_0)\mathrm{cos}(\varphi_0)  \\ (-1)^j \mathrm{sin}(\theta_0)\mathrm{sin}(\varphi_0) \\ \mathrm{cos}(\theta_0)
\end{pmatrix}
,
\label{eq:frustrated_spintube_classical_gs_sup1.5}
\end{equation}
where $\mathrm{cos}(\theta_0)=h/(8SJ_d)$ and we choose
$\varphi_0=0$, this freedom of choice reflecting the U(1) degeneracy
of the ground state.

Then we proceed as usual, allowing these angles to fluctuate by small
quantities $\delta\theta_{\alpha,j}$ and $\varphi_{\alpha,j}$ as
$\theta_0 \rightarrow \theta_{\alpha,j}=\theta_0 + \delta
\theta_{\alpha,j}$, $\varphi_{\alpha}^0 \rightarrow
\varphi_{\alpha,j}$, and introducing new variables $\Pi_{\alpha,j}$
conjugates of the $\varphi_{\alpha,j}$'s. Following the same steps as
previously, namely rewriting the Hamiltonian as a function of these
fluctuation variables, expanding up to second order in the fields and
taking the continuum limit, we obtain the action
\begin{widetext}
\begin{equation}
\begin{split}
S[\{\Pi_{\alpha}\},\{\varphi_{\alpha}\}]=\int d\tau dx \bigg\{&\sum_{\alpha=1,2,3}\left[aJ_{d}(S^2-m^2)(\partial_x\varphi_{\alpha})^2 + a(2J_{d}-J_{\perp})\frac{S^2}{S^2-m^2}\Pi_{\alpha}^2 \right] \bigg.\\
&+\frac{1}{2}(2J_{d}-J_{\perp})\frac{S^2-m^2}{a}\left[(\varphi_1-\varphi_2)^2+(\varphi_2-\varphi_3)^2+(\varphi_3-\varphi_1)^2\right]\\
&+a\left(2J_d\left(1-\frac{m^2}{S^2-m^2}\right)+J_{\perp}\frac{S^2}{S^2-m^2}\right)(\Pi_1\Pi_2+\Pi_2\Pi_3+\Pi_3\Pi_1)\\
&\bigg. +i\sum_{\alpha=1,2,3}\left[\left(\frac{S-m}{a}\right)\partial_{\tau}\varphi_{\alpha}-\Pi_{\alpha}\partial_{\tau}\varphi_{\alpha}\right] \bigg\}.
\end{split}
\label{eq:twisted_spintube_low-energy_action_sup1.5}
\end{equation}
\end{widetext}
The condition $2J_d>J_{\perp}$ for the action to be positive-definite is automatically fullfiled as we
started from the assumption $J_d/J_{\perp}>3/2$. The colinear nature of the classical ground 
state is reflected in the absence of the terms $\Pi_{\alpha}(\varphi_{\alpha+1}-\varphi_{\alpha-1})$ 
(see (\ref{eq:simple_spintube_low-energy_action}) or (\ref{eq:twisted_spintube_low-energy_action_inf1.5})). 

The next steps are to use again the transformation (\ref{eq:orthogonal_transformation_phi}),
perform the gaussian integration in the $\Pi_{\alpha}$ fields and 
rescale the symmetric field as $\phi_s \rightarrow
\phi_s/\sqrt{3}$. This leads to the action
\begin{widetext}
\begin{equation}
\begin{split}
S[\{\phi_{\alpha}\}]&=S_{ch}[\phi_1,\phi_2]+S_s[\phi_s]\\
S[\{\phi_{\alpha}\}]&=\int d\tau dx \bigg\{ \frac{1}{2}\lambda_{\tau}^{(1,2)}\left[(\partial_{\tau}\phi_1)^2+(\partial_{\tau}\phi_2)^2\right]+ \frac{1}{2}\lambda_x^{(1,2)}\left[(\partial_x\phi_1)^2+(\partial_x\phi_2)^2\right]+M^2(\phi_1^2+\phi_2^2) \bigg\}\\
S_s[\phi_s]&=\int d\tau dx \bigg\{ \frac{1}{2}\lambda_{\tau}^{(s)}(\partial_{\tau}\phi_s)^2+\frac{1}{2}\lambda_{x}^{(s)}(\partial_{x}\phi_s)^2+i3\frac{S-m}{a}\partial_{\tau}\phi_s \bigg\}.
\end{split}
\label{eq:twisted_spintube_action_pi_integrated_sup1.5}
\end{equation}
\end{widetext}

Although we end with the same decoupling as previously, this action is actually 
quite different than (\ref{eq:simple_spintube_lowenergy_action_pi_integrated}) and 
(\ref{eq:twisted_spintube_action_pi_integrated_inf1.5}). The crucial point is that 
there is no Berry term $i(\phi_1\partial_{\tau}\phi_2-\phi_2\partial_{\tau}\phi_1)$ 
here. Thus those fields are automatically gapped, and there is no possibility of 
neither an emergent gapless phase nor a spin-imbalance phase (no $i(\Pi\Psi^*-\Pi^*\Psi)$ term), 
contrary to the regime $J_d/J_{\perp}<3/2$ or for 
the simple tube. This is not surprising however, given the fact that 
we are not in the strong-coupling regime and so there is no possibility for the chirality 
described by those two fields to be gapless (or more explicitly, there is no chirality 
in a colinear configuration). Then they only correspond to 
high-energy excitations and we can integrate them out by using the saddle 
point solution $\phi_1=\phi_2=0$.

It remains only the action for the symmetric field, which is exactly 
the same we have already encoutered. We apply the duality transformation 
and repeat our analysis of the dual action. At the end, we recover the 
plateaux condition $3(S-m)\in\mathbb{Z}$. On the other hand, the situation 
is very different for the scaling dimension, as we
find it to be independent of the parameters of the microscopic
model in this second order calculation. It reads
\begin{equation}
\Delta=\frac{3}{2}\pi\sqrt{S^2-m^2}\ .
\label{eq:twisted_spintube_cosine_scalingdim_sup1.5}
\end{equation}
This means that a plateau will either be always present or always
absent when $J_d/J_{\perp}>3/2$. More precisely, if a
plateau is absent for a given spin value, then
it will also be absent for higher spins.

\subsubsection{Discussion}\label{sec:twisted_spintube_path_discussion}
Using the results derived above, we are able to discuss the case of $\mathrm{[(CuCl_2tachH)_3Cl]Cl_2}$,
which belongs to the $J_d/J_{\perp}>3/2$ regime. For $S=1/2$, the OYA
condition predicts only one plateau at a magnetization per spin $m=1/6$,
or $1/3$ of the saturation value. The scaling dimension is
$\Delta=\pi/\sqrt{2}>2$ for these values, so the cosine is
irrelevant. With this result, we predict that the compound does not
possess any plateau in its magnetization curve. In a previous work,
Fouet {\it et al.}~\cite{Fouet2006} reached the same conclusion. They found that
for the realistic coupling values $J_d/J_{\perp}=2.16$, the model
(\ref{eq:frustrated_spintube_Hamiltonian}) behaves as an effective
spin-$3/2$ antiferromagnetic chain with no plateau, and their DMRG
calculations confirmed this effective Hamiltonian approach.

Overall, our analysis is consistent with their numerical phase
diagram based on the DMRG. Indeed, they observed a finite size for the
$m=1/6$ plateau from $J_d/J_{\perp}=0$ to about $J_d/J_{\perp}=3/2$,
where it vanishes. We find here the same result. Altough the form of
the action of the symmetric field is the same starting either from the
$2\pi/3$ state for $J_d/J_{\perp}<3/2$ or from the colinear N\'eel state for
$J_d/J_{\perp}>3/2$, the scaling dimension of the cosine term
indicates that in the first case the plateau is always present while
it disappears in the second one. Thus, we expect qualitatively the
same phase diagram with two different regimes. However,
we should consider carefully this result, as for higher spin $S=3/2$
DMRG calculations seems to indicate that the plateau $m=3/6$ vanishes
also at $J_d/J_{\perp}=3/2$, or very close to this point (see below). Thus the
important point is that we have obtained two different results for the cosine
dimension, one being independent of the microscopic parameters. The
other one tells us that we could observe plateaux depending on the
value $J_d/J_{\perp}$. But we should keep in mind that the critical values
predicted come from a large $S$ analysis.

\subsection{Strongly coupled chains}\label{sec:twisted_spintube_eff_chir}
The triangular unit cell being the same as in the simple tube case,  
the same procedure as done in Sec.~\ref{sec:simple_spintube_eff_chir} can be applied 
to build an effective Hamiltonian on 
the lowest and the highest magnetization plateaux. For instance, the effective 
Hamiltonian to first order perturbation in $J_d/J_{\perp}$ for the spin-1/2 
twisted spin tube on the unique $m=1/6$ plateau reads~\cite{Fouet2006}
\begin{equation}
H_{eff}=\frac{J_{d}}{6}\sum_{j} [1+2 (\tau ^{+}_{j}\tau^{-}_{j+1}+ \tau ^{-}_{j}\tau^{+}_{j+1})],
\label{eq:twisted_spintube_Heff}
\end{equation}
where the chirality operator $\tau$ is defined in (\ref{eq:chirality_operators}). Thus, 
to go beyond this perturbation theory, we propose the same form (\ref{eq:simple_spintube_Heff_generic}) 
for general Hamiltonians describing the emerging chirality degree of freedom along with its bosonized 
form (\ref{eq:simple_spintube_bosonized_effective_Hamiltonian}). It follows that the same succesion 
of a gapless phase then two gapped phases is predicted, as for the simple tube.

Keeping only the XXZ part of (\ref{eq:simple_spintube_Heff_generic}), 
we use again a range-2 CORE calculation as in Sec.~\ref{sec:simple_spintube_range_2_CORE}
to extract numerical values for $J_{xy}$ and $J_z$. Surprisingly, at this level of 
approximation, we find a completely different result compared to the simple tube case,
namely the absence of the XY-ferrochiral transition. For the twisted tube, the 
chirality remains in the $|\Delta|<1$ phase on both plateaux as the coupling $J_{\parallel}$ increases. 
Moreover, we also find this absence of a gapped phase for higher 
half-integer spins, contrary of the simple tube study which shows the 
disappearance of the XY phase in the large-$S$ limit. From this CORE calculation we do no find a 
gapped phase.

\subsection{DMRG results for $S=3/2$}\label{sec:DMRG_twisted_spintube}
\begin{figure}[!htb]
\includegraphics[width=0.45\textwidth,clip]{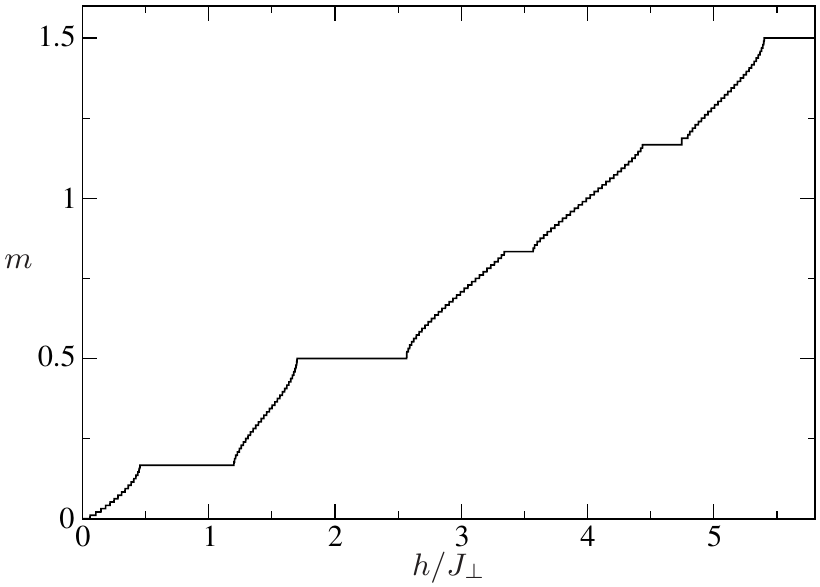}
\caption{Magnetization curve ($m$) vs magnetic field
  $h$ for the twisted three-leg spin tube in the case $S=3/2$. DMRG simulations were 
performed with $L=32$ and $J_d/J_\perp=0.1$}
\label{fig:magnetization_twisted_spintube_S32}
\end{figure}

We now consider the $S=3/2$ case using large-scale numerical
simulations with DMRG algorithm (details are identical to
Sec.~\ref{sec:DMRG_simple_spintube}). First, we confirm that
magnetization plateaux that satisfy OYA criterion
$3(S-m)\in\mathbb{Z}$ exist for small inter-triangle coupling, such as
$J_d/J_\perp=0.1$ as shown in Fig.~\ref{fig:magnetization_twisted_spintube_S32}.

\begin{figure}[t!]
\includegraphics[width=0.45\textwidth,clip]{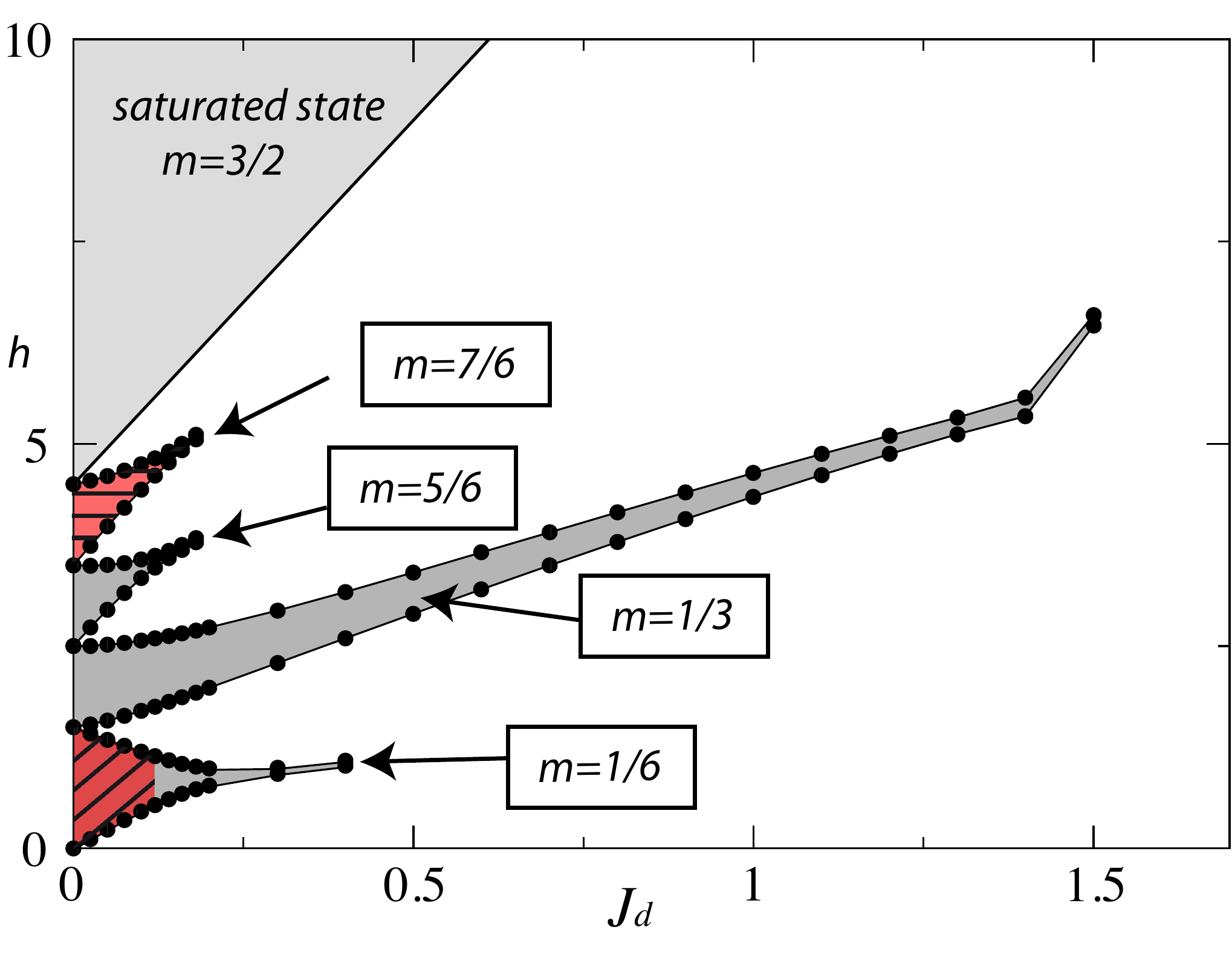}
\caption{(Color online) Phase diagram of the twisted spin tube with $S=3/2$ as 
a function of the couplings $J_d$ and magnetic field $h$ ($J_\perp$ is fixed to 1). 
Several magnetization plateaux can be observed (filled regions) and an additional 
$m=0$ plateau is also found (but is much smaller and not shown on the figure).
 Data correspond to numerical simulations on 
$3\times 32$ lattice with DMRG. Red hashed areas correspond to critical chirality regions (see text).}
\label{fig:phasediag_vs_Jd}
\end{figure}

By performing a finite-size analysis of the plateaux widths, we can obtain 
the phase diagram in magnetic field, shown in Fig.~\ref{fig:phasediag_vs_Jd}. 
As for the simple tube, the precise location of the plateau regions can be 
quite hard due to BKT transitions. In the plot, we have used as a simple criterion 
that the extrapolated plateau should be larger than $0.005 J_\perp$ to be 
considered as finite. Similarly to Fig.~\ref{fig:simple_spintube_phase_diag_S32}, 
there also exists a plateau at $m=0$, which corresponds to a spontaneous dimerization 
of the tube and is not the subject of our present study. We find that the largest 
plateau corresponds to $m=1/3$ and is stable in all the region $J_d/J_\perp\leq 3/2$; 
it exhibits a small anomaly at the tip, similar to what is found in the $S=1/2$ case.~\cite{Fouet2006} 
For a given $J_d/J_\perp$ and increasing $h$, we also note that the order of disappearance 
of the plateaux disagrees with the path-integral prediction, as in the simple 
tube case, which may be due to renormalization effects in the field-theory parameters 
since the field-theory is valid at large $S$ and we are considering $S=3/2$ here. 

\begin{figure}[!htb]
\includegraphics[width=0.45\textwidth,clip]{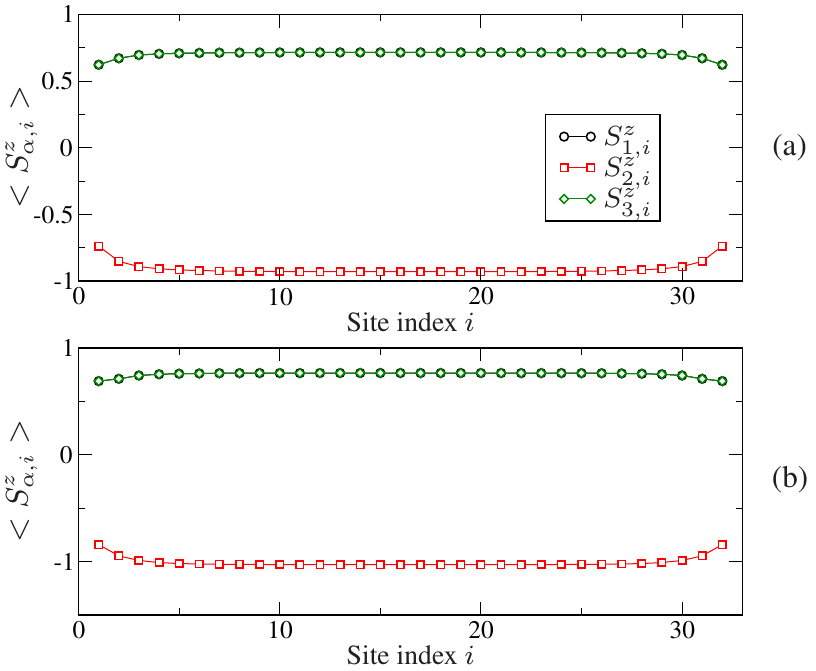}
\caption{(Color online) Local magnetizations obtained by DMRG simulations on a $3\times 32$ twisted spin tube with spin 3/2 on the $m=1/6$ plateau. Upper and lower panels correspond respectively to (a) $J_d/J_\perp=0.2$ and (b) $J_d/J_\perp=0.3$.}
\label{fig:spin_imb_Jd}
\end{figure}

Following Sec.~\ref{sec:simple_spintube_DMRG_entropies}, we now turn to the investigation of the chirality 
for the extreme plateaux $m=1/6$ and $m=7/6$. According to the range-2 CORE Hamiltonian
of the previous section, we expect to observe criticality for these  degrees 
of freedom in all the plateau. Computing the scaling of the block entanglement entropy 
 with DMRG (data not shown), we observe that while it seems to be the case for $m=7/6$, 
we do find a transition to a fully gapped regime in the lower plateau $m=1/6$ when 
$J_d/J_\perp=0.12$ (red hashed regions in Fig.~\ref{fig:phasediag_vs_Jd}). But it turns out 
that the mechanism gapping the chirality is here very different from the scenario determined 
for the simple tube. Computing the local magnetizations for the three chains, we find neither 
a staggered spin-imbalance phase nor a ferrochiral phase. Instead, we find for $J_d/J_{\perp}>0.12$ 
a unique \emph{uniform} spin-imbalance following the critical phase. The imbalance is very strong as shown in 
Fig.~\ref{fig:spin_imb_Jd}, one chain 
having a negative magnetization, and it holds up until the disappearance of 
the plateau. 

The explanation for  the absence of the two gapped phases present in the strong coupling 
limit is given by the reduced density matrix weights. 
In contrast to the simple tube case, the weights of the two chiral states 
rapidy becomes quite small (for instance $47\%$ for a $3\times 6$ 
twisted tube with $J_d/J_\perp=0.3$ as found by ED). Thus, for such values we can not 
rely on the effective Hamiltonian. Note that a similar 
argument involving a mixing with other triangle states, thus prohibiting the 
use of the effective model, was given in Ref.~\onlinecite{Okunishi2012} to 
explain the uniform spin imbalance phase found for the simple $S=1/2$ spin 
tube at magnetization $m=1/3$. Instead we have to rely on the 
path-integral results, valid for any $J_d/J_{\perp}$, which indeed predict 
this uniform spin-imbalance phase. The important remark here is that, 
despite the important difference between the two tubes, we 
are able to understand all the phases observed by combining the results of the path-integral and of 
the bosonized form of the strong-coupling Hamiltonian.

%


\section{Conclusion}\label{sec:conclusion}
In this paper we have studied the magnetic and non-magnetic properties of frustrated 
three-leg spin tubes under a magnetic field. We have considered two kinds of 
geometries, one of which is relevant for the recently studied compound 
$\mathrm{[(CuCl_2tachH)_3Cl]Cl_2}$. Our first result concerns the presence of 
plateaux in the magnetization curve. We give the value of the magnetization at 
which such plateaux can appear given the magnitude of the spins $S$, as well as 
the critical coupling for which such plateaux are expected to appear.
We have used two complementary techniques. The first one is the path integral method 
which, because of the topological nature of the Berry phase term, gives trustable 
qualitative results for any value of $S$, and whose prediction of critical couplings 
are expected to be also quantitatively accurate for large $S$. The second technique 
used is the DMRG method which is however more suited to relatively small spins $S$. 

While magnetization plateaux are not specific to frustrated systems, there 
are emergent low-energy degrees of freedom which presence is due to frustration. 
For historical reasons, we have dubbed those degrees of freedom chirality degrees 
of freedom. Their origin comes from degeneracies in the ground state for 
decoupled triangles, which motivates the use of a third technique to complement 
the path integral and DMRG. This strong coupling technique corresponds to studying 
an effective Hamiltonian in a reduced Hilbert space were high energy degrees 
of freedom are neglected. This supplementary degree of freedom can remain gapless 
even when the magnetization degrees of freedom are gapped (in the magnetization plateau). 
Here again, the agreement between the three techniques for predicting the critical 
couplings at which such degree of freedom disappear is qualitatively excellent and 
quantitatively quite satisfactory. It is important to stress that, although we restricted 
ourselves to the cases of two degenerate spin-$1/2$ representations for half-integer spin 
cases, the chirality is in principle a generic feature that can give rise to more 
complicated effective Hamiltonians. It could arise also for integer spin tubes, 
provided they are tuned to the appropriate value of the magnetization.

The results obtained here and the excellent complementarity of the path-integral 
technique, the effective hamiltonian approach and the DMRG calculations are very 
encouraging. Indeed, the study of potentially gapless non-magnetic degrees of 
freedom has become a central topic in the study of exotic phases in frustrated 
quantum magnetism. In the systems analyzed here, one important generalization 
that deserves futures studies is the interplay of doping with such degrees of freedom. 
It is by now relatively well established that doping will result in shifts and 
splitting of the magnetization plateaux in such quasi one-dimensional systems \cite{Cabra2001,Cabra2002,Lamas2011}, 
but how doping may affect the non-magnetic degrees of freedom is still an open problem. 
The other important extension concerns the role of such non-magnetic degrees of 
freedom in higher dimensional frustrated systems, where some results for distorted 
kagome lattices are indeed quite instructive.~\cite{Subrahmanyam1995,Mila1998} A chirality also appears 
for instance in the study of trimerized Mott insulators.~\cite{Kamiya2012} The emergence of 
new analytical and numerical techniques for studying such issue in two and three 
dimensional frustrated magnets are also very promising.



\acknowledgments

We would like to thank P. Lecheminant, E. Orignac and K. Totsuka for enlightening discussions. 
Numerical simulations were performed using HPC resources from GENCI-IDRIS (Grant 2012050225) and CALMIP.

\appendix

\section{Duality transformation}\label{sec:appendix}

We present here the details of the duality transformation used on the
symmetric part $S_s$ of the action (\ref{eq:simple_spintube_lowenergy_action_pi_integrated}). 
First, we perform a Hubbard-Stratonovich transformation introducing an
auxiliary field $\vec{J}=(J_{\tau},J_x)$ and we divide the field
$\phi_s$ in two parts $\phi_s=\phi_{s,v} + \phi_{s,f}$, where
$\phi_{s,f}$ has no vorticity
$(\partial_{\mu}\partial_{\nu}-\partial_{\nu}\partial_{\mu})\phi_{s,f}=0$. 
The action reads
\begin{equation}
\begin{split}
S[\phi_{s,v},\phi_{s,f},\vec{J}]&=\int d\tau dx \bigg\{ \frac{1}{2\lambda_{\tau}^{(s)}}J_{\tau}^2 + \frac{1}{2\lambda_{x}^{(s)}}J_{x}^2 \bigg.\\
&+ i\left(J_{\tau}+3\frac{S-m}{a}\right)\partial_{\tau}\phi_{s,v} + iJ_x\partial_x\phi_{s,v} \\
&+i\left(J_{\tau}+3\frac{S-m}{a}\right)\partial_{\tau}\phi_{s,f} + iJ_x\partial_x\phi_{s,f} \bigg\}.
\end{split}
\label{eq:appendix_duality_transformation_1}
\end{equation}
Integrating by parts the last two terms containing the vorticy-free component, 
the action takes the form
\begin{equation}
\begin{split}
S[\phi_{s,v},\phi_{s,f},\tilde{\vec{J}}]&=\int d\tau dx \bigg\{ \frac{1}{2\lambda_{\tau}^{(s)}}\left(\tilde{J}_{\tau}-3\frac{S-m}{a}\right)^2 \bigg.\\
&+ \frac{1}{2\lambda_{x}^{(s)}}\tilde{J}_{x}^2 + i(\tilde{J}_{\tau}\partial_{\tau} + \tilde{J}_x\partial_x)\phi_{s,v} \\
&- i(\partial_{\tau}\tilde{J}_{\tau} + \partial_x\tilde{J}_x )\phi_{s,f}\bigg\},
\end{split}
\label{eq:appendix_duality_transformation_2}
\end{equation}
where we have defined
$\tilde{\vec{J}}=(J_{\tau}+3\frac{S-m}{a},J_x)$. The vorticity-free
part simply leads to a zero divergence constraint on the auxiliary
field $\partial_{\tau}\tilde{J}_{\tau} + \partial_x\tilde{J}_x =0$ and
we obtain
\begin{equation}
\begin{split}
S[\phi_{s,v},\tilde{\vec{J}}]=\int d\tau dx \bigg\{ &\frac{1}{2\lambda_{\tau}^{(s)}}\left(\tilde{J}_{\tau}-3\frac{S-m}{a}\right)^2 \bigg.\\
&+ \frac{1}{2\lambda_{x}^{(s)}}\tilde{J}_{x}^2 + i\tilde{J}_{\mu}\partial_{\mu}\phi_{s,v}\bigg\}.
\end{split}
\label{eq:appendix_duality_transformation_3}
\end{equation}
The constraint can be solved in one dimension by introducing the dual
field $\Phi_s$ defined by 
$\tilde{J_{\mu}}=\epsilon_{\mu\nu}\partial_{\nu}\Phi_s$, this field beeing
vorticity-free. Then we integrate by parts the last term in
(\ref{eq:appendix_duality_transformation_3}) and, with the redefinition
$\tilde{\Phi}_s=\Phi_s-3\frac{S-m}{a}x$, we get
\begin{equation}
\begin{split}
S[\tilde{\Phi}_s]=\int d\tau dx \bigg\{ &\frac{1}{2\lambda_x^{(s)}}(\partial_{\tau}\tilde{\Phi}_s)^2 + \frac{1}{2\lambda_{\tau}^{(s)}}(\partial_x\tilde{\Phi}_s)^2 \bigg.\\
&+ i2\pi\rho_v\left(\tilde{\Phi}_s+3\frac{S-m}{a}x\right)\bigg\}.
\end{split}
\label{eq:appendix_duality_transformation_4}
\end{equation}
In this action $\rho_v$ is the space-time density of vortices defined
as
$(\partial_{\tau}\partial_{x}-\partial_{x}\partial_{\tau})\phi_v=\epsilon_{\mu\nu}\partial_{\mu}\partial_{\nu}
\phi_v=2\pi\sum_jq_{j,v}\delta(\tau-\tau_{j,v})\delta(x-x_{j,v})=2\pi\rho_v$
with $(\tau_{j,v},x_{j,v})$ the space-time coordinates of the $j$-th
vortex and $q_{j,v}\in\mathbb{Z}$ its charge. After summing over all
the vortex configurations in the partition function and rescaling the
imaginary time, we finally end with the action
\begin{equation}
\begin{split}
S[\tilde{\Phi}_s]=\int d\tau dx \bigg\{ &\frac{1}{2}K(\vec{\nabla}\tilde{\Phi}_s)^2 \bigg.\\
&+ g_1\mathrm{cos}\left(2\pi\left[\tilde{\Phi}_s+3\frac{S-m}{a}x \right]\right) \bigg\},
\end{split}
\label{eq:appendix_duality_transformation_final}
\end{equation}
where $K=1/\sqrt{\lambda_{\tau}^{(s)}\lambda_x^{(s)}}$, $\vec{\nabla}=(\partial_{\tau},\partial_x)$ and $g_1$ is a constant we have not calculated.

%


\end{document}